\documentclass[preprint,12pt,times]{myelsarticle}

\pdfoutput=1

\usepackage[colorlinks]{hyperref}  
\hypersetup{pdfpagemode=None,pdfstartview=FitH}
\usepackage{fancyheadings}
\usepackage{fullpage}
\usepackage{amsmath}
\usepackage{amsbsy}
\usepackage{amssymb}
\usepackage{amsthm}
\usepackage{amscd}
\usepackage{amsfonts}
\usepackage{supertabular}
\usepackage{graphics}
\usepackage{graphicx}
\usepackage{verbatim}
\usepackage{subcaption}
\usepackage{epsfig}
\usepackage{xspace}
\usepackage{multirow}
\usepackage{euscript}
\usepackage{alltt}
\usepackage{boxedminipage}
\usepackage{float}
\usepackage{dsfont}
\usepackage{mathrsfs} 
\usepackage{color}
\usepackage{bm}
\usepackage{natbib}

\usepackage{pgfplots}
\usepackage{tikz}
\usetikzlibrary{shapes,calc,arrows,through,intersections}

\newcommand{\vm}[1]{\bm{#1}}
\newcommand{\vx}{\vm{x}}

\newcommand{\xv}{\vx}
\newcommand{\tauv}{\vm{\tau}}

\newcommand{\Rv}{\mathbf{R}}
\newcommand{\av}{\vm{a}}
\newcommand{\kv}{\vm{k}}

\newcommand{\calW}{{\mathcal{W}}}
\newcommand{\del}{\nabla}

\newcommand{\eps}{\varepsilon}
\newcommand{\nhat}{\hat{\vm{n}}}
\newcommand{\Vloc}{{V^{\ell}}}
\newcommand{\Vnloc}{{\hat{V}^{n\ell}}}

\newcommand{\eref}[1]{(\ref{#1})}
\newcommand{\fref}[1]{Fig.~\ref{#1}}
\newcommand{\tref}[1]{Table~\ref{#1}}
\newcommand{\sref}[1]{Section~\ref{#1}}
\newcommand{\cref}[1]{Chapter~\ref{#1}}

\renewcommand{\rm}[1]{\textrm{#1}}
\renewcommand{\it}[1]{\textit{#1}}

\renewcommand{\Re}{{\mathbb{R}}}

\journal{{\textrm Extreme Mechanics Letters}}

\graphicspath{ {./} {./figs/} }

\biboptions{sort&compress}

\begin{document}

\begin{frontmatter}

\title{Partition of unity finite element method for quantum mechanical materials calculations}

\author[label1]{J. E. Pask\corref{cor1}}
\ead{pask1@llnl.gov}
\author[label2]{N. Sukumar\corref{cor1}}
\ead{nsukumar@ucdavis.edu}
\cortext[cor1]{Corresponding authors}

\address[label1]{Lawrence Livermore National Laboratory,
                 Livermore, CA 94550, USA.}
\address[label2]{Department of Civil and Environmental Engineering, 
                 University of California, Davis, CA 95616, USA.}

\begin{abstract}
The current state of the art for large-scale quantum-mechanical simulations is the 
planewave (PW) pseudopotential method, as implemented in codes such as
VASP, ABINIT, and many others.
However, since the PW method uses a global Fourier
basis, with strictly uniform resolution at all points in space, it suffers from substantial 
inefficiencies in calculations involving atoms with localized states, such as first-row and 
transition-metal atoms, and requires significant nonlocal
communications, which limit parallel efficiency. 
Real-space
methods such as finite-differences (FD) and finite-elements (FE) have partially addressed both resolution 
and parallel-communications issues but have been plagued by one key
disadvantage relative to PW: excessive number of degrees of freedom (basis functions) 
needed to achieve the required accuracies. In this paper, we present 
a real-space partition of unity finite element (PUFE) method to solve the
Kohn-Sham equations of density functional theory. In the PUFE method, we build the known 
atomic physics into the solution process using 
partition-of-unity enrichment techniques in finite element
analysis. The method developed herein is completely general, 
applicable to metals and insulators alike, 
and particularly efficient for deep, localized potentials, as occur in calculations at extreme conditions of pressure and temperature.  
Full self-consistent Kohn-Sham calculations
are presented for LiH, involving light atoms, and  
CeAl, involving heavy atoms with large numbers of 
atomic-orbital enrichments. We find that the new PUFE approach attains the required
accuracies with substantially fewer degrees of
freedom, typically by an order of magnitude or more, than the PW method. 
We compute the equation of state of LiH
and show that the computed
lattice constant and bulk modulus are 
in excellent agreement with
reference PW results, while requiring an order of magnitude fewer degrees of freedom to obtain.
\end{abstract}


\end{frontmatter}

\section{Introduction}
First principles ({\em ab initio\/}) quantum mechanical simulations 
based on density functional theory (DFT)~\cite{HohK64,KohS65} are a 
vital component of research in condensed matter physics and molecular quantum 
chemistry. The parameter free, quantum-mechanical nature of the
theory facilitates both fundamental understanding and robust
predictions across the gamut of materials systems, from metallic actinides
to insulating organics, at ambient and extreme conditions alike.
However, the solution of the equations
of DFT is a formidable task and this has severely limited the range of 
materials systems that can be investigated by such accurate 
means.  First and foremost, an \it{ab initio} quantum mechanical 
description is required whenever departures from
isolated-atomic or known condensed-matter configurations may be
significant, as in materials at extreme conditions of pressure and temperature 
where charge is transferred, new phases emerge, and bonds are formed and broken.  
However, a merely \it{ab initio} 
approach is not enough
in the investigation of such unfamiliar systems: the
approach must be general, equally applicable to all atomic species and
configurations, and systematically improvable so that errors can be
clearly known and strictly controlled. 

For condensed matter systems---solids, liquids, and mixed-phase---the 
planewave (PW) pseudopotential method~\cite{Pic89} is among the most 
widely used \it{ab initio} methods that afford this level of 
generality and systematic improvability.  The accuracy and 
generality of the PW method arises from its nature and basis: a 
variational expansion approach in which solutions are represented in a 
Fourier basis. By virtue of the completeness of the basis, any 
condensed matter system can be modeled with arbitrary accuracy, in 
principle, by simply adding sufficient wavenumbers to the basis. In 
practice, however, the PW method has significant limitations 
with respect to the solution of large, complex problems: 
a global Fourier basis with 
uniform resolution at all points in space, 
leading to inefficiencies in calculations with localized states 
and significant nonlocal communications which compromise parallel efficiency.
These limitations have 
inspired extensive research over the past two decades into {\em real-space methods},
among the more mature of which are the finite-difference (FD),
finite-element (FE), and wavelet based approaches~\cite{Ar99,Bec00,TorEE06,PasS05b,GenNG08,Saad:2010:NMF}. 

The advantages of a strictly local, real-space approach in large-scale 
calculations have been amply demonstrated in the context of finite-difference 
(FD) methods~\cite{ChelTS94,SeitPN95,GygG95,IyMB95,HosAF95,BrigSB96,ModZK97,Fat99,Bec00,FB:1,FG:3,AlJK04,TorEE06,Beck:2009:RSM,Ghosh:2016:SPARC1,Ghosh:2016:SPARC2}. 
These methods allow for some variable resolution in real space, can 
accommodate a variety of boundary conditions, and require no 
computation- or communication-intensive transforms. Finite-difference
methods achieve these advantages, however, by giving up the use of a basis 
altogether, instead
discretizing operators directly on a real-space grid, which leads to
disadvantages such as limited accuracy in
integrations~\cite{BrigSB96,On99,TorEE06} and nonvariational
convergence~\cite{Bec00,TorEE06}. Furthermore,
because FD methods lack a basis, it is difficult to build known physics
into these methods in order to increase the efficiency of the representation.  
Despite these disadvantages, however, it has recently been shown that finite-difference codes 
can outperform established planewave codes not only in
isolated-system (e.g., molecule, cluster) calculations but in condensed matter (e.g., solid, liquid)
calculations as well~\cite{Ghosh:2016:SPARC1,Ghosh:2016:SPARC2} in
moderately large calculations on parallel computers. Such FD codes 
may thus present a viable alternative to planewave codes in such a context.

Applications of the finite element method~\cite{StranF73} to 
the electronic structure of atoms and molecules go back to the
1970s~\cite{As75,WhitWT89} (see, e.g., Ref.~\cite{Beck:2009:RSM} for a review).
Applications to condensed matter systems appeared about a
decade later~\cite{HerY86,TsucT95a,TsucT98,PasS05b}, and 
have been in active development by a number of groups since 
then~\cite{TsucT95a,TsucT96,TsucT98,PasKF99,PasKS01,PasS05a,PasS05b,TorEE06,Motamarri:2013:HOA,Tsuc15}. 
In the present work, we shall focus on condensed matter systems;
however, the methodology presented is equally applicable to isolated systems as well.
Early solid-state calculations~\cite{TsucT95a,TsucT96,TsucT98,PasKF99,PasKS01} employed 
relatively low-order elements, typically cubic or lower.
Due to the relatively high accuracy required in quantum mechanical calculations, however, 
it was soon appreciated that 
higher-order elements should be 
advantageous~\cite{Bat00,LehHP09,Motamarri:2013:HOA} to reduce 
the degrees of freedom (DOFs) (basis functions) needed to attain the 
required accuracies. It
has recently been shown that such a high-order spectral-element (SE)
formulation can outperform established planewave codes in calculations
of isolated systems, though not yet for condensed matter systems~\cite{Motamarri:2013:HOA}. 
While the number of DOFs required can be greatly reduced, it still
exceeds that required by standard planewave methods in
the context of such condensed matter systems. Moreover, as polynomial order
is increased in SE methods, the cost per DOF increases. 
However, superior
parallelization of SE methods relative to PW 
stands to compensate these disadvantages in large-scale calculations as SE based methods are further developed.

The significant flexibility of the finite element method to
concentrate degrees of freedom in real space where needed and omit
them where not, via standard $h$- or $p$-adaptive refinement 
techniques~\cite{Bat00,Zhang:2008:FEM,BylHW09,LehHP09,Alizadegan:2010:ADA,Bao:2012:HAF,Schauer:2013:AEK,Schauer:2014:RBM,Motamarri:2013:HOA,Motamarri:2014:SSS,Chen:2014:AFE,Maday:2014:HPF,Tsuc15,Davydov:2016:AFE}
allows a much more efficient representation of 
highly inhomogeneous problems, as occur in \it{ab initio} electronic 
structure, than is possible by either FD or PW approaches. By 
exploiting only the \it{scale} of variations and not the 
known orbital \it{nature} in electronic structure, however,
such methods must invariably overcome a quite substantial DOF
disadvantage relative to mature orbital-based methods~\cite{Skriv84,SinN06,Boys:1950:EWF}
in order to be competitive, in addition to overcoming the overhead of
mesh generation and associated load balance issues in parallel implementations.  
To overcome the critical DOF disadvantage of FE methods relative to
PW, the known atomic physics must be leveraged. The Kohn-Sham
wavefunctions in a multi-atom system are well approximated by
isolated-atom wavefunctions near the atom centers. And isolated-atom
wavefunctions are readily computed using standard radial
solvers. Furthermore, it is in precisely this region, near the atom
centers, where the multi-atom system wavefunctions vary most rapidly.
Hence, a substantial reduction in DOFs stands to be achieved by adding 
isolated-atom wavefunctions to the standard FE basis, thus leaving the FE basis
to take up only the smooth perturbation away from isolated atom solutions, as 
the atoms are brought together to form a molecule or solid. 
Dusterhoft and coworkers~\cite{DusHK98} exploited this idea 
by adding atomic orbitals to a Lagrange
FE basis in 2D axisymmetric all-electron calculations of C$_2$, whereas
Yamaka and Hyodo~\cite{YamH03,YamH05} 
demonstrated the substantial gains in efficiency that can be realized 
by adding well chosen Gaussian basis functions to the FE basis in
 full 3D all-electron molecular calculations. 
A significant disadvantage of 
the above approaches, however, as noted in Ref.~\cite{YamH05}, is 
that by adding such extended functions to the otherwise strictly local
FE basis, sparseness and locality are compromised, leading to
inefficiencies in parallel implementation. 

Modern partition of unity finite element (PUFE) methods
provide an elegant and highly efficient solution to 
the quantum-mechanical problem. 
Jun~\cite{Jun04} has reported non-self-consistent results using a meshfree 
basis to solve the Schr{\"o}dinger equation in periodic solids, 
whereas Chen and coworkers~\cite{Che08} and Suryanarayana and 
coworkers~\cite{Surya:2011:AMC} have employed meshfree partition-of-unity basis functions 
in full self-consistent calculations. 
Finite element basis functions 
as the partition-of-unity, however, offer significant advantages: 
locality is retained to the maximum 
extent possible, which facilitates parallel implementation, 
variational convergence is ensured 
by the min-max theorem~\cite{StranF73}, 
numerical integration errors can be strictly controlled, 
and Dirichlet as well as periodic boundary conditions are readily 
imposed.  Sukumar and Pask~\cite{Sukumar:2009:CEF,Pask:2012:LSS} have
demonstrated the substantial reductions in DOFs that can be achieved
by PUFE relative to standard FE in solutions of Schr{\"o}dinger and
Poisson equations in both pseudopotential and all-electron
calculations, and have shown those advantages are retained in
Kohn-Sham calculations~\cite{Pask:2011:PUF}.

In this work, we show the substantial degree-of-freedom
reduction afforded by the PUFE solution of the Kohn-Sham equations relative
to both classical FE and current state-of-the-art PW methods. In particular,
we show in calculations of total energy, equation of state, lattice
constant, and bulk moduli that the PUFE method can achieve an
order-of-magnitude or more reduction in DOFs relative to
state-of-the-art PW methods, while remaining systematically improvable
and strictly local in real space, thus facilitating efficient parallel implementation. 

The remainder of this paper is organized as follows. 
In~\sref{sec:pufem}, we introduce the essentials of the partition of unity finite element method 
and discuss its particular suitability for quantum mechanical problems.
In \sref{sec:ksdft}, we review the Kohn-Sham equations to be solved, and 
in \sref{sec:sol} discuss solution in a PUFE basis.
We discuss key implementational issues in \sref{sec:imple} and 
present applications to systems involving both light and heavy atoms in \sref{sec:results}, 
where we find order-of-magnitude reductions in basis size relative to 
current state-of-the-art methods in calculations of total energy, equation of state, lattice constant, and bulk modulus.
Finally, we conclude with our main findings and future outlook
in~\sref{sec:conclusions}.

\section{Partition-of-unity finite elements}\label{sec:pufem}
The partition of unity finite element method~\cite{Melenk:1996:PUF,Babuska:1997:PUM}
generalized the standard finite element method by providing a
means to incorporate analytical as well as numerical solutions of boundary-value
problems into the finite element approximation.  Any set of functions
$\{ \phi_i(\vx) \}_{i=1}^n$ that sum to unity is said to form a
 \it{partition-of-unity} (PU).
For $C^0$ Lagrange finite elements, a basis function $\phi_j(\vx)$ 
is associated with node $j$ in the mesh. 
Consider a domain $\Omega \subset \Re^d$.
The PUFE approximation 
for a scalar-valued function $u:\Omega \rightarrow \Re$ 
is of the form~\cite{Melenk:1996:PUF}:
\begin{equation}\label{eq:pum}
u^h(\vx) = \sum_{i=1}^n \phi_i(\vx) u_i + \sum_{\alpha=1}^m
\sum_{j=1}^{\bar n} \phi_j^\textrm{PU}(\vx) \Psi_\alpha(\vx) b_{j\alpha} ,
\end{equation}
where $\phi_i(\vx)$ are FE basis functions, $\phi_j^\text{PU}(\vx)$ form a partition of unity,
and $\Psi_\alpha(\vx)$ are referred to as \it{enrichment functions};  so that
$u_i$ are the standard finite element degrees of freedom,
and ${b}_{j\alpha}$ are additional enriched degrees of freedom. 
In the present quantum mechanical context, we construct efficient enrichment functions $\Psi_\alpha(\vx)$ from isolated-atom solutions and employ trilinear FE basis functions $\phi_j^\textrm{PU}(\vx)$ for the partition of unity in order to impose strict locality while minimizing additional DOFs. Moreover, we employ higher-order FE basis functions $\phi_i(\vx)$ to efficiently attain required accuracies. An illustration of the substantial gains afforded by such enrichment is provided in the Supplementary Materials. We elaborate on the application to the Kohn-Sham equations below.

\section{Kohn-Sham equations}\label{sec:ksdft}
Kohn-Sham density functional theory (KS-DFT) 
replaces the $N$-body problem of interacting electrons
by a computationally tractable problem of noninteracting
electrons moving in an effective mean field. To this end,
one-electron Schr\"odinger equations for Kohn-Sham orbitals
are solved, which are used to predict the ground-state electronic 
density, total energy, and related material properties. The equations of DFT 
require the repeated solution of a one-electron Schr{\"o}dinger 
equation in the presence of an effective potential that consists of 
nuclear (external potential), Hartree (electron-electron repulsion), 
and exchange-correlation contributions. In KS-DFT, the 
exchange-correlation functional is not known exactly 
and so approximations must be made. Common approximations include the local density approximation 
(LDA)~\cite{KohS65} and generalized gradient approximation (GGA)~\cite{perdew:1996:GGA}, 
the former of which we employ in the present work.

The Schr\"odinger equation for the $i$-th KS-orbital is:
\begin{equation}\label{eq:KS}
{\cal H}\psi_i := 
\left[ -\dfrac{\nabla^2 \psi_i(\vx)}{2} + \hat{V}_{\rm{eff}}(\vx) 
\right] \psi_i(\vx) = \eps_i \psi_i(\vx),
\end{equation}
where ${\cal H}$ is the Hamiltonian operator, 
$\psi_i$ is the $i$-th wavefunction, 
$\eps_i$ is the corresponding energy eigenvalue, 
and we have used atomic units here and throughout. 
The effective potential
\begin{equation}
\hat{V}_{\rm{eff}}(\vx) = V_n(\vx) + V_H(\vx;\rho_e(\vx)) + V_{xc}\bigl(\vx;\rho_e(\vx)\bigr),
\end{equation}
where $V_n(\vx)$ and $V_H(\vx;\rho_e(\vx))$ are the nuclear potential and
the Hartree potential, respectively, 
$V_{xc}(\vx;\rho_e(\vx))$ is the exchange correlation potential, and
$\rho_e(\vx)$ is the electronic charge density.
From the KS-orbitals, the electronic charge density is constructed as
\begin{equation}\label{eq:rhoe}
\rho_e = - \sum_{k=1}^{N_{\rm{occ}}} f_k|\psi_k|^2 ,
\end{equation}
where $ 0 \le f_k \le 2$ specifies the occupation of state $k$ 
(neglecting spin) and $N_{\rm{occ}}$ is the number of occupied states.

\begin{figure}
\centering
\includegraphics[width=0.69\textwidth]{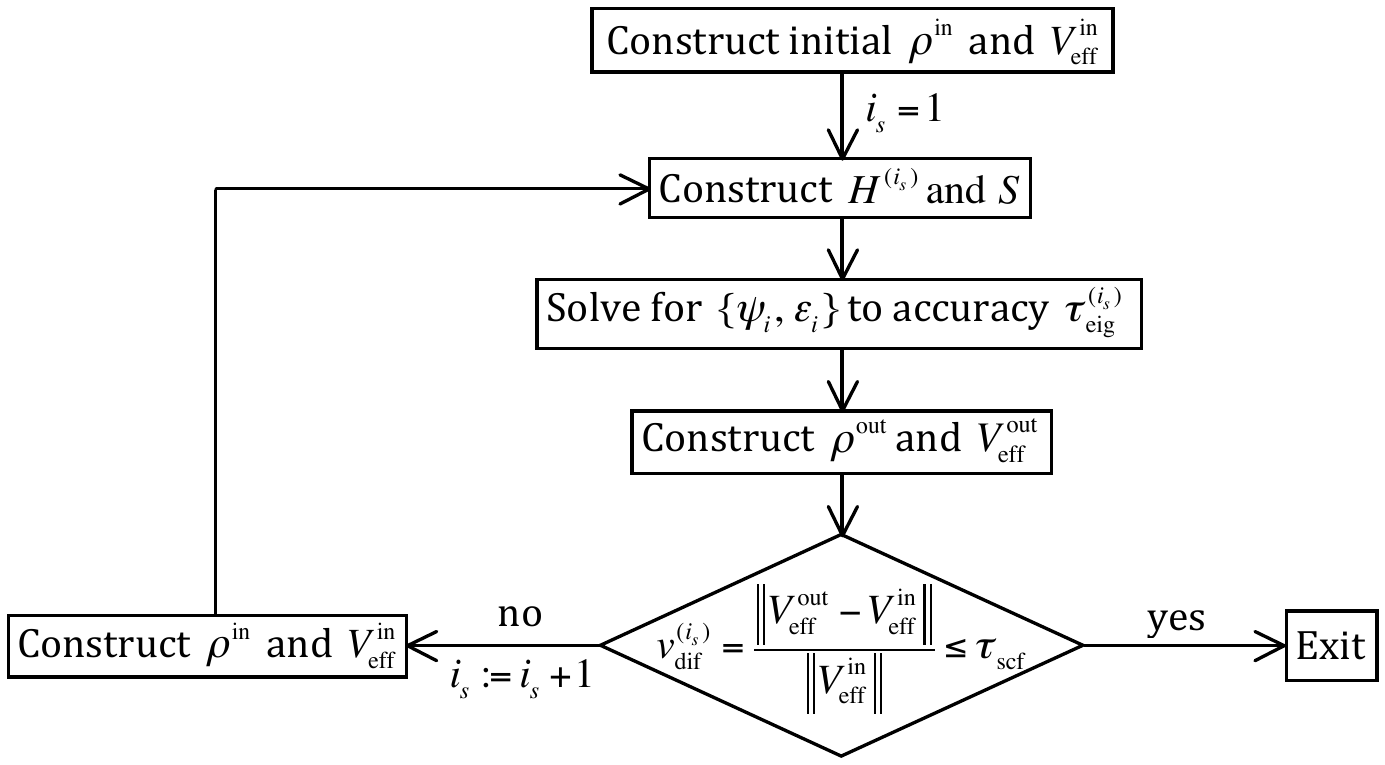}
\caption{Self-consistent solution procedure.}\label{fig:scf}
\end{figure}

In the pseudopotential approximation~\cite{Pic89}, 
the potential of the nucleus and core
electrons is replaced by a smooth, effective \emph{pseudopotential}, 
expressed as a sum of local ionic ($V_I^{\ell}$)
and nonlocal ionic ($\hat{V}_I^{n\ell}$) contributions. 
This leaves only smooth valence states to be determined, 
which are readily computed by PW and real-space methods.
The Kohn-Sham effective potential is then given by~\cite{Pic89,PasS05a}:
\begin{subequations}\label{eq:veff}
\begin{align}
\label{eq:veff-a}
\hat{V}_{\rm{eff}}(\vx) &= V_I^\ell(\vx) + \hat{V}_I^{n\ell}(\vx)
+ V_H(\vx;\rho_e) + V_{xc}(\vx;\rho_e), \\
\label{eq:veff-b}
V_I^\ell(\vx) &= \sum_a V_{I,a} (\vx), \\
\label{eq:veff-c}
\hat{V}_I^{n\ell}(\vx) \psi_i &= \sum_a \int V_{I,a}^{n\ell} (\vx,\vx^\prime)
\psi_i(\vx^\prime) \, d\vx^\prime, \\
\label{eq:veff-d}
V_H(\vx;\rho_e) &= -\int \dfrac{\rho_e(\vx^\prime)}{|\vx - \vx^\prime|} d\vx^\prime,
\end{align}
\end{subequations}
where $V_{I,a}$ and $V_{I,a}^{n \ell}$ are the local and nonlocal parts of the
ionic pseudopotential of atom $a$, the integrals extend over all space, and
the summations over all atoms $a$ in the system. 
Since the ionic potentials $V_{I,a}$ decay as $Z_a/r$, with $Z_a$ the valence charge of atom $a$, 
they can be generated by ionic charge densities $\rho_{I,a}$ 
strictly localized in real space. The total charge density can then be constructed as 
$\rho = \rho_I + \rho_e$, where $\rho_I = \sum_a \rho_{I,a}$~\cite{PasS05a}.
Instead of the integral \eref{eq:veff-d} for the Hartree term, 
we can then compute the total Coulomb potential $V_C = V_I^\ell + V_H$ 
by solving the Poisson equation 
\begin{equation}\label{eq:poisson}
\nabla^2 V_{C} = 4 \pi \rho,
\end{equation}
whereupon \eref{eq:veff-a} can be reduced to
\begin{equation}\label{eq:veff-localnonlocal}
\hat{V}_{\rm{eff}}(\vx) =
\bigl( V_C(\vx;\rho_e) + V_{xc}(\vx;\rho_e) \bigr) 
+ \hat{V}_I^{n\ell}(\vx) 
\equiv V^\ell (\vx) + \hat{V}^{n \ell} (\vx).
\end{equation}
Since $\psi_i$ in \eref{eq:KS} depends on $\hat{V}_{\rm{eff}}$, which depends on $\rho$, 
which itself depends on $\psi_i$ in \eref{eq:rhoe}, the KS equations constitute a
nonlinear eigenvalue problem.
The equations are typically solved by 
fixed-point (``self-consistent field'') iteration~(\fref{fig:scf}), 
whereby the unique energy-minimizing density $\rho_e$ is obtained, corresponding to 
the physical electronic density in the molecular or condensed matter system. 
Energies, forces, and other properties of interest are then computed from the 
self-consistent density and eigenfunctions so obtained~\cite{Mar04}.
In the present work, we compute total energies as in Pask and Sterne~\cite{PasS05a}.
\section{Solution in PUFE basis}\label{sec:sol}
The key computational step in the solution of the Kohn-Sham equations is the solution of the effective Schr\"odinger equation \eref{eq:KS} for all occupied eigenfunctions $\psi_i(\vx)$ and associated eigenvalues $\eps_i$. We now focus on this key step in the context of condensed matter calculations.

We consider a parallelepiped unit-cell domain $\Omega \subset \Re^3$ defined by 
primitive lattice vectors $\av_i$ ($i=1,2,3$).
In a periodic system, the charge density and electrostatic potential are periodic, i.e.,
\begin{subequations}
\begin{gather}
\label{eq:periodicrho}
\rho(\xv + \Rv) = \rho(\xv), \\
\label{eq:periodicV}
V(\xv + \Rv) = V(\xv),
\end{gather}
\end{subequations}
and the wavefunction $\psi$, the solution of Schr\"odinger's 
equation, satisfies Bloch's theorem
\begin{equation}\label{eq:bloch}
\psi(\xv + \Rv) = \exp(i \kv \cdot \Rv)\psi(\xv),
\end{equation}
where $\Rv = i_1\av_1 + i_2\av_2+ i_3\av_3$ ($i_1,\,i_2,\,i_3 \in \mathbb{Z}$) is
a lattice translation vector, $\kv$ is the wavevector, and 
$i = \sqrt{-\!1}$~\cite{ashcroft:book}. For $\kv = \vm{0}$ ($\Gamma$-point), 
the wavefunction is periodic; otherwise, there is a phase shift $\exp(i\kv \cdot \Rv)$ with translation $\Rv$ from cell to cell in the periodic system. 

The strong form of the Schr\"odinger problem \eref{eq:KS} in the unit cell is then~\cite{PasKF99}:
\begin{subequations}
\label{eq:strong}
\begin{gather}
\label{eq:schr}
-\tfrac{1}{2}\del^2 \psi(\xv) + \Vloc(\xv) \psi(\xv) 
+ \ \Vnloc (\vx) \psi(\xv) = \eps \psi(\xv) \quad \textrm{in } \Omega, \\
\label{eq:bbc1}
\psi(\xv + \av_\ell) = \exp(i \kv \cdot \av_\ell) \psi(\xv) \quad \textrm{on } \Gamma_\ell, \\
\label{eq:bbc2}
\del \psi(\xv + \av_\ell) \cdot \nhat = \exp(i \kv \cdot \av_\ell) \del \psi(\xv) \cdot \nhat \quad \textrm{on } \Gamma_\ell,
\end{gather}
\end{subequations}
where $\psi$ is the wavefunction (eigenfunction), $\Vloc$ and 
$\Vnloc$ are the local and nonlocal potentials defined in \eref{eq:veff-localnonlocal},
$\eps$ is the energy eigenvalue, $\av_\ell$ are the primitive lattice vectors, 
$\nhat$ is the unit outward normal at $\xv$, and $\Omega$ and 
$\Gamma_\ell$ are the domain and bounding surfaces, respectively.
Note that, since the
boundary conditions are complex-valued, 
the wavefunctions are in general complex also; 
however, due to the self-adjoint operator, the eigenvalues are real.

Now, since we wish to discretize in a $C^0$ basis, we require the weak form of \eref{eq:strong}.
On taking the inner product of the differential 
equation~\eqref{eq:schr} with an arbitrary (complex-valued)
test function $w(\xv)$, applying the divergence theorem, and
imposing the derivative boundary condition~\eref{eq:bbc2}, we 
obtain the weak form~\cite{PasS05b}: 
Find functions $\psi \in \calW$ and scalars $\eps \in \Re$ such that
\begin{subequations}
\label{eq:schrweakform}
\begin{gather}
\label{eq:schrweak}
a(w,\psi) = \eps (w,\psi) \quad \forall w \in \calW, \\
a(w,\psi) = \dfrac{1}{2} \int_\Omega \del w^* \cdot \del \psi \, d\xv 
+ \int_\Omega w^* \Vloc \psi \, d\xv  
+ \int_\Omega w^* \Vnloc \psi \, d\xv , \quad
(w,\psi) = \int_\Omega w^* \psi \, d\xv , \\
\label{eq:wbpspace}
\calW = \Bigl\{ w \in H^1(\Omega): 
w(\xv+\av_\ell) = \exp(i \kv \cdot \av_\ell) w(\xv) \textrm{ on } 
\Gamma_\ell \Bigr\}.
\end{gather}
\end{subequations}

We now discretize the weak form in the PUFE basis
\begin{equation}\label{eq:pufebasis}
\bigl\{ \varphi_k(\xv) \bigl\} \ \equiv \bigl\{ \phi_i(\xv) \bigr\} \cup 
\bigl\{ \phi_j^\text{PU}(\xv)\Psi_\alpha(\xv) \bigr\},
\end{equation}
where $\phi_i(\xv)$ are the standard FE basis functions, 
$\phi_j^\text{PU}(\xv)$ form the partition of unity, and 
$\Psi_\alpha(\xv)$ are the enrichment functions.
The trial wavefunction in the PUFE method is then~\cite{Sukumar:2009:CEF}:
\begin{equation}\label{eq:psih}
\psi^h(\xv) = \sum_{i \in \mathds{I}} \phi_i(\xv) u_i 
+ \sum_\alpha
\sum_{j \in \mathds{J}} \phi_j^\textrm{PU}(\xv) \Psi_\alpha(\xv) 
b_{j\alpha} \equiv \sum_{k=1}^{N} \varphi_k(\xv) c_k ,
\end{equation}
where $\mathds{I}$ is the index set 
consisting of all nodes in the mesh, $\mathds{J} \subseteq \mathds{I}$, 
$N$ is the total number of basis functions (degrees of freedom), 
and $u_i$ and $b_{j\alpha}$ are nodal coefficients
associated with the FE and PU enrichment bases, respectively.
Substituting the trial function~\eref{eq:psih} into the weak 
form~\eref{eq:schrweakform} 
and taking $\varphi_i(\xv)$ as test functions then
yields a sparse Hermitian generalized eigenproblem for the eigenvalues
and eigenfunction coefficients~\cite{PasS05b,Sukumar:2009:CEF}:
\begin{subequations}\label{eq:disceig3d}
\begin{align}
\label{eq:Hc}
\vm{H}\vm{c} &= \eps \vm{S}\vm{c}, \quad \vm{c} = \{\vm{u} \ \vm{b}\}^T, \\
\vm{H}_{ij} &= \int_\Omega \Bigl( \tfrac{1}{2} \del \varphi_i^\ast \cdot
\del \varphi_j + \varphi_i^\ast \Vloc \varphi_j 
            + \varphi_i^\ast \Vnloc \varphi_j \Bigr) \, d\vx, \\
\vm{S}_{ij} &= \int_\Omega \varphi_i^\ast \varphi_j \, d\vx, \\
\intertext{where for a separable nonlocal potential~\cite{Pic89}, 
the nonlocal term is evaluated as~\cite{PasS05b}:}
\int_\Omega \varphi_i^\ast \Vnloc \varphi_j \, d\vx &=
\sum_{a,L} f_{L}^{ai} h_L^a \left(f_L^{aj}\right)^\ast, \quad
f_{L}^{ai} = \int_\Omega \varphi_i^\ast(\vx) \sum_n
e^{i \kv \cdot \vm{a}_n} v_L^a (\vx - \tauv_a - \vm{a}_n ) \, d\vx,
\end{align}
\end{subequations}
where 
$a$ and $L$ are atom- and angular momentum indices, respectively, 
$h_L^a$ are weights, $v_L^a(\vx - \tauv_a - \vm{a}_n)$ 
are projector functions associated with image atoms at $\tauv_a+\vm{a}_n$, 
and $n$ runs over all lattice vectors.

Solution of the discrete eigenproblem \eref{eq:Hc} for eigenvalues and 
eigenfunction coefficients then yields the PUFE approximations 
$\eps^h$ and $\psi^h(\xv) = \sum_k \varphi_k(\xv) c_k$ of the 
eigenvalues $\eps$ and eigenfunctions $\psi$ of the 
continuous eigenproblem \eref{eq:strong}; 
which for self-consistent Kohn-Sham potential $\Vloc$ 
are the desired Kohn-Sham eigenvalues and eigenfunctions 
from which materials properties are computed.

\section{Implementation}\label{sec:imple}
\subsection{Partition of unity enrichment functions}
The key to the efficiency of the PUFE method is the accuracy with which enrichment functions $\Psi_\alpha(\xv)$ added to the standard FE basis $\{\phi_i(\xv)\}$ represent the solutions sought, 
in the present case, the Kohn-Sham wavefunctions in the condensed matter system of interest. To construct such accurate enrichment in the quantum mechanical context, we exploit the fact that the wavefunctions of a condensed matter system resemble those of the constituent atoms in the vicinity of each atom; the closer to the atom center, the closer the resemblance. Hence, isolated-atom wavefunctions stand to be an excellent choice for enrichment. However, the atomic wavefunctions have extended exponential tails, which would require significant storage and long-range lattice summations to obtain all contributions in the unit cell. On the other hand, the information of interest in the atomic wavefunctions is contained within a few atomic units of the atom centers, the remainder of the wavefunctions being smoothly decaying and easily represented by the classical FE part of the basis. Therefore, for efficiency, we use smoothly truncated atomic wavefunctions for enrichment rather than the extended atomic wavefunctions themselves.

Since the potential of an isolated atom is spherical, the eigenfunctions have the form
\begin{equation}\label{eq:Psitilde}
\tilde{\Psi}_{n\ell m}(\xv) = R_{n\ell}(r)Y_{\ell m}(\theta,\phi),
\end{equation}
where $Y_{\ell m}(\theta,\phi)$ are the spherical harmonics and 
$R_{n\ell}(r)$ can be obtained by solving
the associated radial Schr{\"o}dinger equation~\cite{Griffiths:book}, 
which we carry out using a high-order spectral solver. 
The numerically computed radial solutions are output at discrete points and
to ensure continuity and value-periodicity~\eref{eq:wbpspace}
of the PUFE approximation, we
set $R_{n \ell}(r)$ to be the cubic spline-fit of the product of
the numerically computed pseudoatomic wavefunction 
$\tilde{R}_{n\ell}(r)$ and
a $C^3$ cutoff function 
\begin{equation*}
h(r,r_c) = 
\begin{cases}
1 + \dfrac{20r^7}{r_c^7} - \dfrac{70r^6}{r_c^6} + 
\dfrac{84 r^5}{r_c^5} - \dfrac{35r^4}{r_c^4}, & r \le r_c \\
0, & r > r_c
\end{cases},
\end{equation*}
where $r_c \equiv r_c^{n \ell}$ is the \emph{cutoff radius}, which limits
the extent of the enrichment function. The smooth cutoff
also ensures that the enrichment function is smooth within each
element in order to facilitate efficient quadrature.
The value of $r_c^{n \ell}$ is set so that second nearest neighbor cells are
sufficient to determine the total enrichment contribution in the unit cell. For enriching
a valence state ($n$ and $\ell$ specified),  all nodes
that lie within the \emph{support radius} $r_e^{n \ell}$ of  
the atom are enriched. Since the enrichment functions are most rapidly
varying at small $r$ and smoothly varying at large $r$, as $r_e^{n \ell}$ approaches
$r_c^{n \ell}$, the condition number of the system matrices worsens, and 
therefore one strikes a compromise between improved accuracy and 
ill-conditioning, which dictates the choice of $r_e^{n \ell}$.
To construct the enrichment function associated with a given atomic state (indexed by quantum numbers $n$, $\ell$, and $m$) in the unit cell,
second-nearest-neighbor cell contributions are summed to 
form
\begin{equation}\label{eq:Psi}
\Psi_{n \ell m} (\xv) = \sum_{\Rv} f(\Rv) \tilde{\Psi}_{n \ell m}(|\xv - \tauv - \Rv|),
\end{equation}
where $f(\Rv) = 1$ for periodic enrichment and 
$f(\Rv) = \exp(i \kv \cdot \Rv)$ for Bloch-periodic enrichment.
The enrichment function is centered at the atom position $\tauv$ and
the lattice translation vectors
\mbox{$\Rv = i_1\vm{a}_1 + i_2\vm{a}_2 + i_3\vm{a}_3$}
($i_i, i_2, i_3 = -2, \ldots, 2$).

Once the enrichment functions $\Psi_\alpha(\xv)$ are constructed, there remains the choice of partition of unity ${\phi_j^\text{PU}(\xv)}$ to form the final, strictly local partition of unity enrichment functions ${\phi_j^\text{PU}(\xv)}\Psi_\alpha(\xv)$ of the PUFE basis. In order to minimize the number of partition of unity functions needed, we choose trilinear FE basis functions for the partition of unity, regardless of polynomial order chosen for the classical FE part of the PUFE basis, which is typically chosen to be higher-order in order to most efficiently attain the accuracies required with minimal degrees of freedom.

\subsection{Adaptive quadrature}
For hexahedral finite elements, it suffices to perform numerical integration
using tensor-product Gauss quadrature rules.
For PUFE, since the enrichment functions adopted to solve the Schr{\"o}dinger
eigenproblem are nonpolynomial with
sharp, local variations, accurate and efficient
integration of the weak form integrals is needed to realize optimal rates
of convergence and ensure satisfaction of the min-max theorem.
Furthermore, since an atom can be located anywhere inside an element, 
many of the standard methods for numerical integration 
prove to be inefficient. To address this, we employ an adaptive
quadrature scheme~\cite{Mousavi:2012:EAI} which attains high accuracy
at minimal cost, and which provides the required efficiency to
solve the quantum-mechanical problem.

Since numerous integrals are carried out during the course of the Kohn-Sham solution---for matrix elements, normalizations, total energies, etc.---and are carried out a number of times as the fixed point iteration proceeds, we construct a single quadrature rule for all integrals based on adaptive quadrature of the most rapidly varying integrands, rather than perform adaptive quadrature afresh for each integral individually. 
To construct a rule sufficient for all integrals, we consider the terms 
$\Psi^2$ and $\del \Psi \cdot \del \Psi$ ($\Psi$ is the enrichment function) 
that appear in the integrands of the Hamiltonian and overlap matrices.
We then construct a quadrature rule over each 
finite element that satisfies a given error tolerance for these
integrands. We begin with a $5 \times 5 \times 5$ tensor-product Gauss
quadrature rule over each element. 
An $8 \times 8 \times 8$ tensor-product Gauss rule is then used 
to compute the reference value for these integrals to estimate the local 
error. If the absolute error of integration is larger than the 
prescribed tolerance, the integration domain is uniformly subdivided into eight 
cells and the adaptive integration is performed over each cell recursively. 
This process is started for the two integrands indicated above, and
at each step only those functions whose integration error is higher than the 
tolerance are passed to the next level. This process is repeated
until all functions are accurately integrated, and the
output is an optimized nonuniform quadrature rule consisting of 
quadrature points and weights in each element. In the numerical computations,
we use a tolerance of $10^{-6}$ to generate the adaptive quadrature
rule over each element. Further details on the adaptive scheme can 
be found in Ref.~\cite{Mousavi:2012:EAI}.
\section{Results and discussion}\label{sec:results}
\subsection{Model pseudopotential}\label{subsec:modelpsp}
\begin{figure}
\centering
  \begin{subfigure}{0.40\textwidth}
     \includegraphics[width=\textwidth]{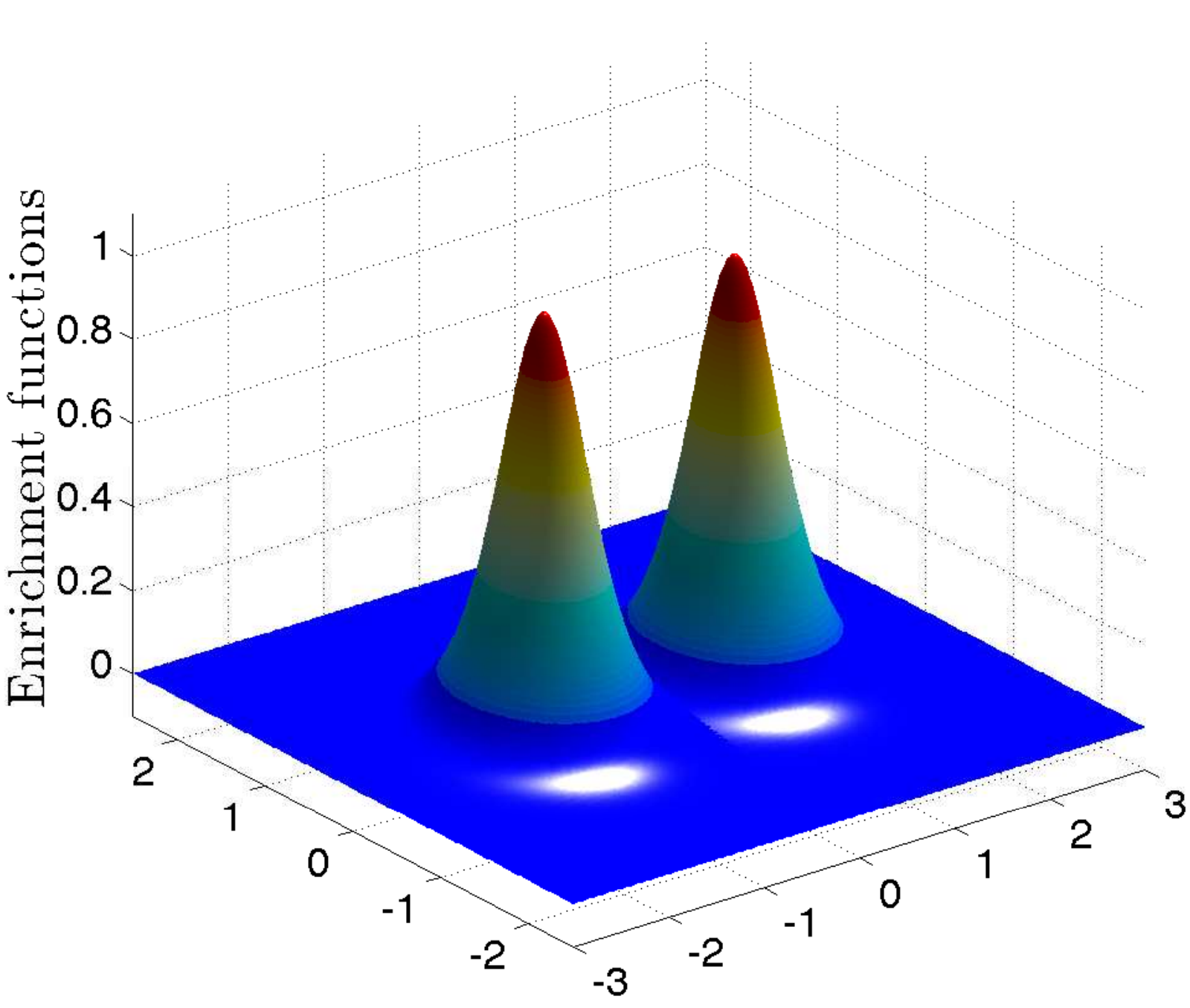}
     \caption{}
     \label{fig:doublewell-a}
  \end{subfigure}
  \qquad 
  \begin{subfigure}{0.48\textwidth}
     \includegraphics[width=\textwidth]{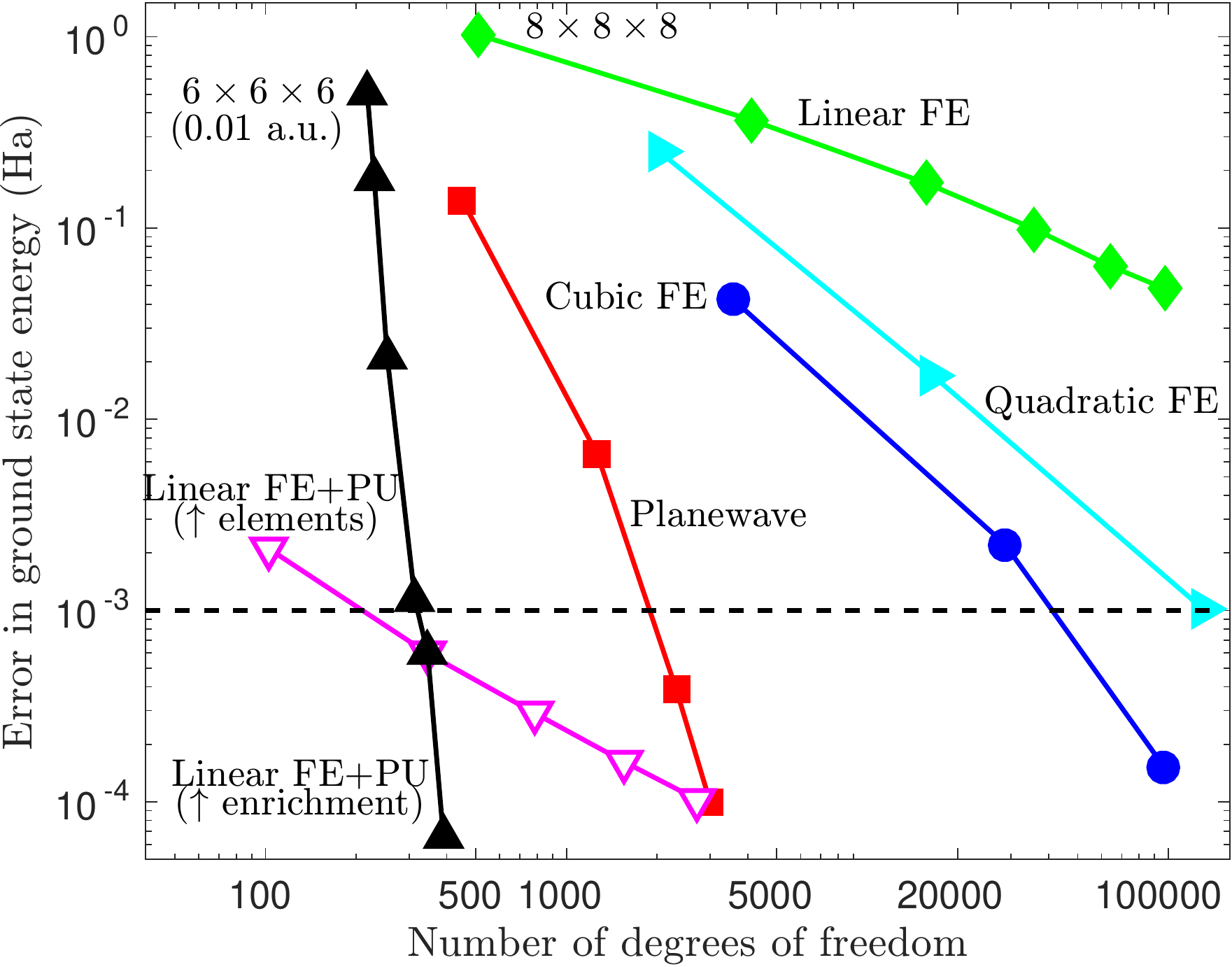}
     \caption{}
     \label{fig:doublewell-b}
  \end{subfigure}
\caption{Localized (model pseudopotential) test problem. 
        (a) Enrichment functions at $z = 0$. 
        (b) Error in computed ground state energy 
         versus number of degrees 
         of freedom using standard PW, real-space FE, and
         real-space PUFE methods. The coarsest mesh for
         linear FE is $8 \times 8 \times 8$. 
         Linear FE + PU ($\uparrow$ enrichment) corresponds to 
         linear FE with $6 \times 6 \times 6$ mesh and enrichment
         support radius increasing from $0.01$ a.u.\ to $2.51$ a.u.
         Linear FE + PU ($\uparrow$ elements) corresponds to 
         linear FE with fixed enrichment support radius of $2.15$ a.u. 
         and increasing number of elements in each direction.}
         \label{fig:doublewell}
\end{figure}

To assess the efficiency of the PUFE method relative to current 
state-of-the-art electronic-structure methods for the Schr\"odinger
eigenproblem, we compare PUFE head-to-head against a standard
PW code and modern conventional FE codes on a standard test 
problem with a localized potential of the form~\cite{Gyg92,TsucT96}:
\begin{equation*}
\Vloc(r) = V_0 \exp\left(-\frac{r^2}{r_0^2}\right), \quad V_0 = -16,
    \  r_0 = 0.5,
\end{equation*}
which is representative of typical {\em ab initio} pseudopotentials for 
elements such as
transition metals and actinides. 
The domain $\Omega = (-3,3)\times (-2.5,2.5) \times
(-2.5,2.5)$, the total potential contribution is the sum of
Gaussian potentials (double-well) that are 2 a.u.\ apart, and
each atom is enriched separately. 
The two enrichment
functions are shown in~\fref{fig:doublewell-a}.
 
Figure~\ref{fig:doublewell-b} shows the error in the computed ground state 
energy for the localized test problem versus number of degrees of 
freedom using standard PW, real-space FE, and real-space PUFE methods. 
The horizontal line indicates an error of $10^{-3}$ Ha, 
as typical in \it{ab initio} calculations. First, we see that that 
PW method requires fewer degrees of freedom to reduce the error
to this level than any of the standard FE methods. 
This demonstrates the significant DOF disadvantage that must be overcome 
by such methods to compete effectively with 
current state-of-the-art PW based methods.  
Next, we consider the linear FE + PU result.
This achieves the required level of accuracy with a factor of 5 fewer 
degrees of freedom than the PW method, using just the simplest, least
accurate uniform-mesh linear FE basis. Quadratic and cubic FE
bases would converge even faster. Even for this moderately
localized potential and linear FE basis, the advantages of 
PU enrichment are substantial.

\subsection{Self-consistent calculations}\label{subsec:scc}
Since the Fourier basis is global, the convergence of energies and 
eigenvalues with respect to number of basis functions (degrees of freedom) 
in the 
PW method is spectral, i.e., faster than any polynomial. In contrast,
since finite element basis functions are polynomial in nature, the convergence of FE and 
other such real-space methods is determined by polynomial completeness. 
For a polynomial basis that is complete to order $p$, the errors in 
energies and eigenvalues of self-adjoint operators are $O(h^{2p})$,
whereas the errors in the associated eigenfunctions are 
$O(h^{p+1})$~\cite{StranF73}. Hence, for sufficiently high accuracies, the 
spectral convergence of PW-based methods dominate and require fewer 
DOFs than fixed-degree polynomial based methods such as finite elements.
However, at lower accuracies, with partition-of-unity enrichment in particular, 
the FE basis can require fewer DOFs, substantially fewer in fact, as we show below. 

In the PUFE calculations that follow, 
we use cubic (serendipity) finite elements with 32 nodes per element,
multiplied by appropriate phase factors at domain boundaries~\cite{Sukumar:2009:CEF},
to form the Bloch-periodic classical FE part $\{\phi_i(\xv)\}$ of the PUFE basis
and trilinear finite elements with 8 nodes, 
multiplied by appropriate phase factors at domain boundaries, 
to form the Bloch-periodic partition of unity $\{\phi_j^\textrm{PU}(\vx)\}$.
With both Bloch-periodic FE and Bloch-periodic partition of unity, 
the enrichment functions $\Psi_\alpha(\xv)$ are then constructed to be 
periodic~\cite{Sukumar:2009:CEF}, with $f(\Rv) = 1$ in~\eref{eq:Psi}.

\subsubsection{LiH}
We compare the PUFE method head-to-head against the current state-of-the-art PW method and 
standard FE method in total energy calculations of LiH, using hard, 
transferable HGH pseudopotentials~\cite{HarGH98}. 
The unit cell is cubic with lattice parameter $a = 4.63$ bohr. The
position of the Li atom in lattice coordinates is $\tauv = \vm{0}$ and
that of the H atom is $\tauv = (1/2,1/2,1/2)$.
 
\begin{figure}
\centering
  \begin{subfigure}{0.40\textwidth}
     \includegraphics[width=\textwidth]{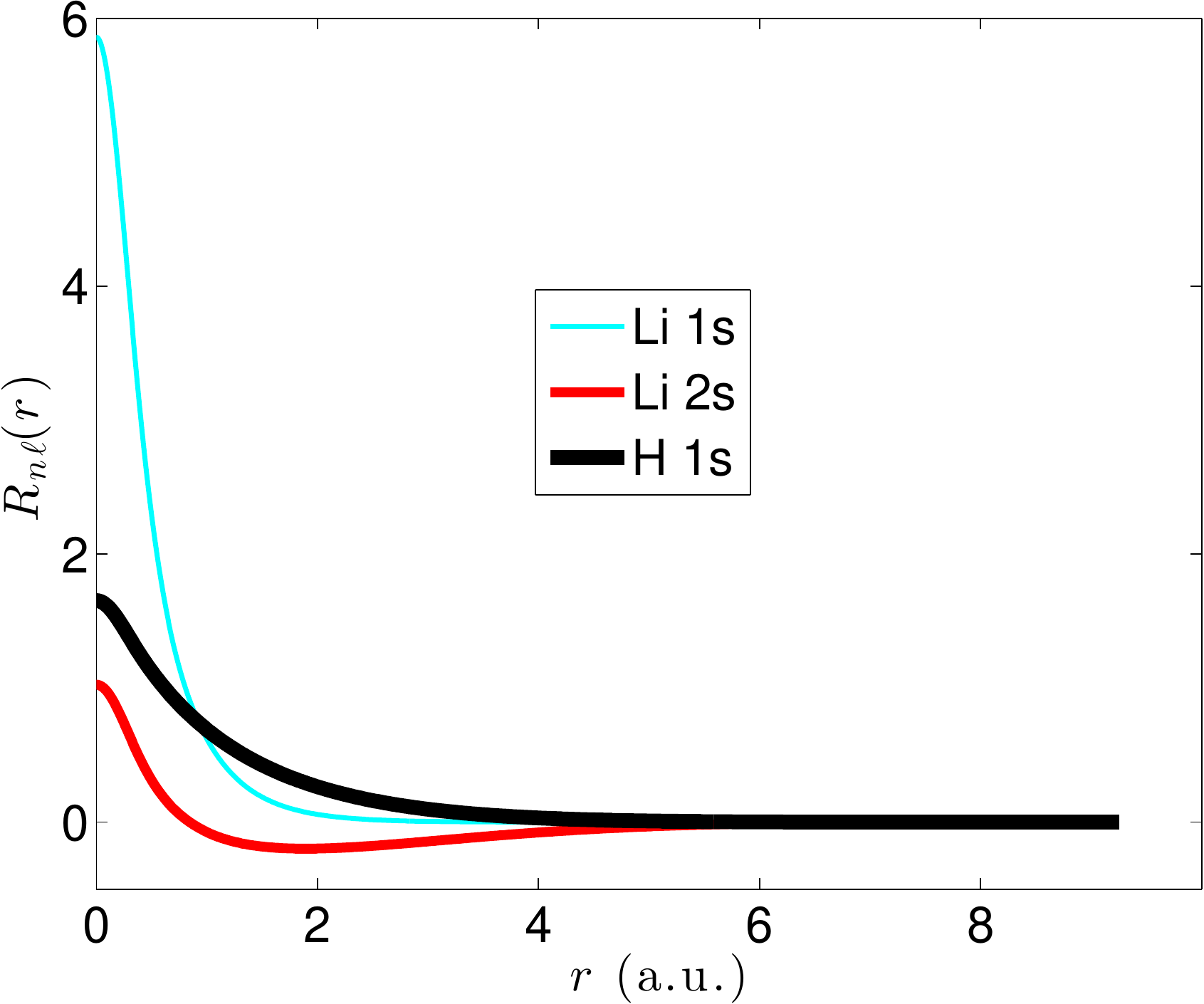}
     \caption{}
     \label{fig:LiH-a}
  \end{subfigure}
  \qquad 
  \begin{subfigure}{0.48\textwidth}
     \includegraphics[width=\textwidth]{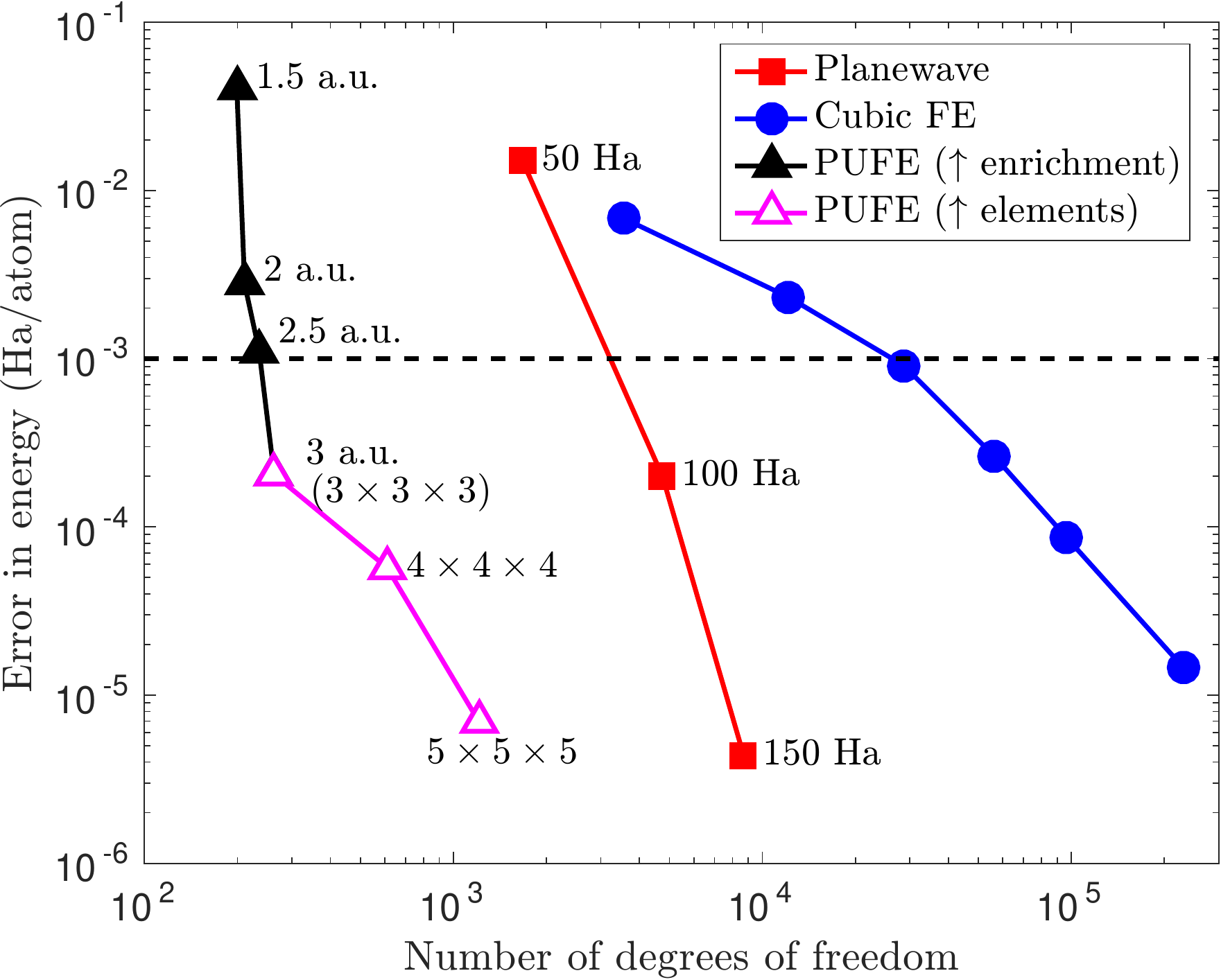}
     \caption{}
     \label{fig:LiH-b}
  \end{subfigure}
\caption{Enrichment functions and convergence of total energy 
         for Kohn-Sham 
         calculations of LiH.
         (a) Radial part of enrichment functions 
         (1s and 2s for Li, 1s for H). 
         (b) Error in computed total energy per atom of LiH versus
         number of degrees of freedom using standard PW, real-space FE, and
         real-space PUFE methods. 
         Enrichment support radii and mesh sizes are indicated next to 
         corresponding data points for PUFE.
         Planewave cutoffs are indicated next to corresponding 
         data points for PW.}
         \label{fig:LiH}
\end{figure}

Two enrichments are used for Li (1s and 2s states in valence) and one
enrichment for H (1s state in valence).  The cutoff radius $r_c^{n\ell}$ 
for all enrichment functions is set to $2a$. 
Plots of the radial parts of the enrichment functions are shown 
in~\fref{fig:LiH-a}. 
The Brillouin zone is sampled at two $\kv$-points:
$\kv = (0.00, \, 0.00,\,  0.00)$ and
$\kv = (0.12, \, -0.24,\, 0.37)$. 
The reference result for the total energy is 
$-8.091160$ Ha,
obtained from a PW calculation with 300 Ha planewave cutoff.

Figure~\ref{fig:LiH-b} shows the error in computed total energy for LiH
versus number of degrees of freedom (basis functions) using PW, 
real-space FE, and real-space PUFE methods. The horizontal
line indicates an error of $10^{-3}$ Ha/atom, as typical in \it{ab initio} 
calculations. 
We first note that PW calculations require a factor of 9 
fewer DOFs to reduce 
the error to the desired level than the real-space cubic FE method, which 
demonstrates the extent of the DOF disadvantage real-space methods must 
overcome to be competitive with planewaves. We now consider the cubic PUFE 
result. This achieves the required level of accuracy with a 
factor of 12 fewer DOFs than the PW method.  
The significant benefits of PU enrichment are manifest. 
On a fixed $3 \times 3 \times 3$ cubic FE mesh,
increasing the enrichment support radii from 1.5 a.u.\ to 3.0 a.u.\ 
(just 64 additional DOFs) renders a more than two order of magnitude improvement
in accuracy. Then, on refining the mesh for a fixed
enrichment support radius of 3 a.u., the convergence rate 
of the PUFE error becomes comparable to standard cubic FE.

While increasing accuracy dramatically and retaining strict locality, 
the improvements afforded by partition-of-unity enrichment 
come with some cost: worsened conditioning. 
To quantify these aspects, 
we compute the sparsity and condition number of the 
overlap matrix in LiH calculations for a series of enrichment support radii. 
In Table~\ref{table:sparsity}, we report the
sparsity and condition numbers of the overlap matrix 
for a $3 \times 3 \times 3 $ cubic FE mesh and 
enrichment support radii 
$r_e = 0, 1, 2, 3$ a.u. ($r_e = 0$ is a standard cubic FE calculation).  
For a fixed mesh spacing $h$, the condition number of the PUFE 
overlap matrix worsens with increasing enrichment support radius.
Condition numbers in the range $10^7$ to $10^9$ 
in Table~\ref{table:sparsity} are comparable to 
those of high-quality Gaussian basis sets in quantum
chemistry~\cite{Rappoport:POG:2010}.
The PUFE solution with $r_e = 3$ 
delivers less than 1 mHa/atom error in energy (\fref{fig:LiH-b}).
From Table~\ref{table:sparsity}, 
the sparsity of the overlap matrix is, however, just $38.3 \, \%$. 
This is a consequence of the small 2-atom unit cell.
Since the PUFE basis is strictly local, the number of nonzeros per row of the overlap matrix 
is independent of system size, and so sparsity increases with systems size. 
For example, for 
a 16-atom unit cell with 
$6 \times 6 \times 6 $ cubic FE mesh and $r_e = 3$, the sparsity
of the overlap matrix increases to $92.1 \, \%$. 
For small problems, where dense matrix methods can be applied, the conditioning presents no particular problem. However, for larger problems, where iterative methods must be employed, the conditioning requires special attention, as discussed in Ref.~\cite{Cai:2013:HPI}.

\begin{table}
\caption{Sparsity and condition number of PUFE overlap matrix
         for LiH calculation.} 
\label{table:sparsity}
\medskip
\centering
\begin{tabular}{c c c c}
\hline
\hline
Method & $r_e$ & Sparsity (\%) & Condition number \\ 
\hline
FE     &   --   &  51.7  &  $2.2 \times 10^3$  \\ 
PUFE   &   1.0   &  51.2  &  $9.2 \times 10^6$  \\ 
PUFE   &   2.0   &  46.8  &  $1.3 \times 10^8$  \\ 
PUFE   &   3.0   &  38.3  &  $2.3 \times 10^9$  \\ 
\hline
\hline
\end{tabular}
\end{table}

\subsubsection{CeAl}
To assess the performance of the PUFE method in the worst case, we apply it 
to a difficult f-electron system: triclinic CeAl using hard, 
nonlocal HGH pseudopotentials~\cite{HarGH98}, with atoms
displaced from ideal positions.  The unit cell is triclinic, with 
primitive lattice vectors and atomic positions
\begin{subequations}\label{eq:ceal-primitive}
\begin{align}
\av_1 &= a(1.00 \quad 0.02 \quad -0.04),\nonumber \\
\av_2 &= a(0.01 \quad 0.98 \quad  0.03),\\
\av_3 &= a(0.03 \quad -0.06 \quad 1.09),\nonumber \\
\vm{\tau}_{\rm{Ce}} &= a(0.01 \quad 0.02 \quad 0.03), \\
\vm{\tau}_{\rm{Al}} &= a(0.51 \quad 0.47 \quad 0.55),
\end{align}
\end{subequations}
with lattice parameter $a = 5.75$ bohr and atomic positions in lattice
coordinates.

\begin{figure}
\centering
   \begin{subfigure}{0.35\textwidth}
      \includegraphics[width=\textwidth]{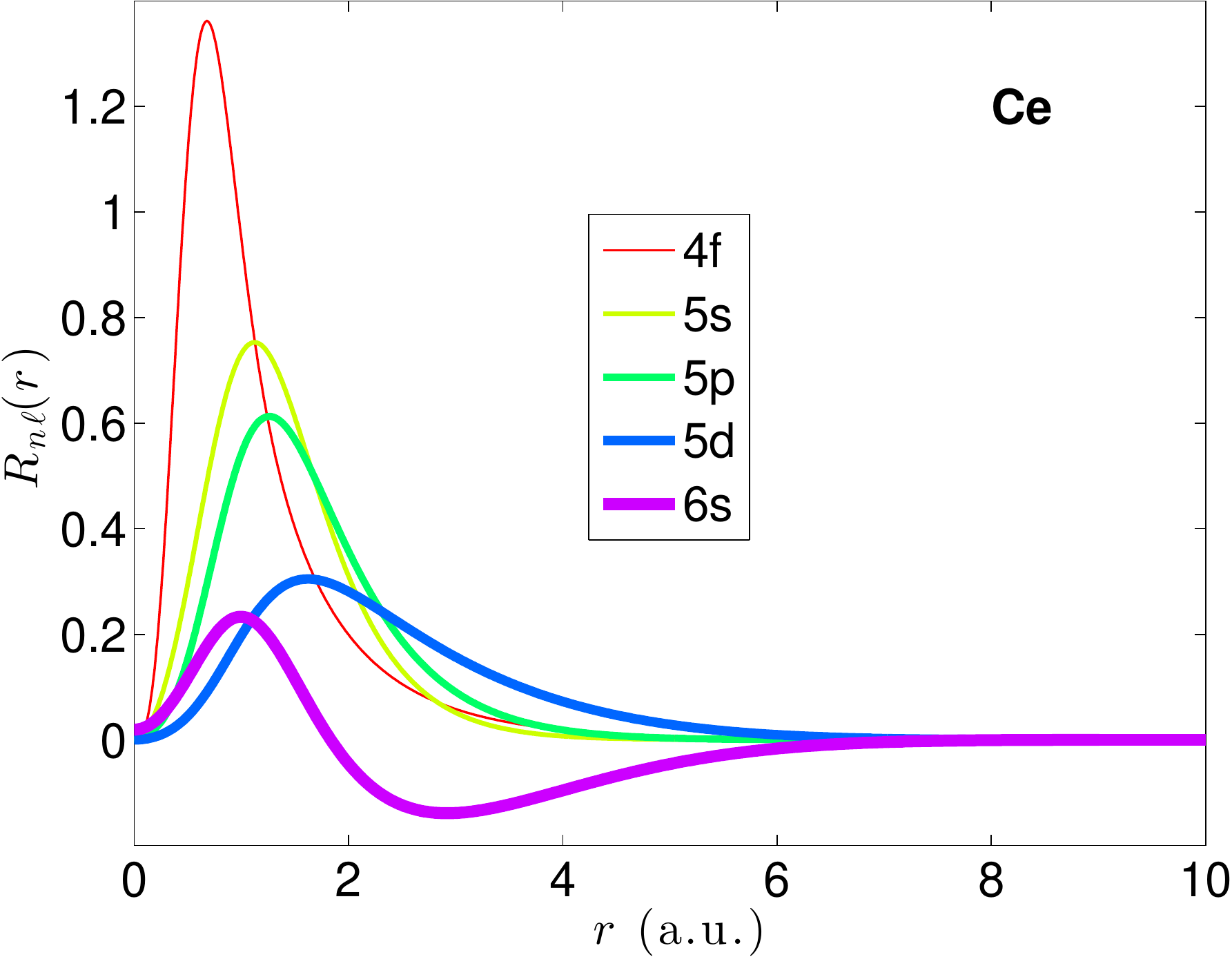}
      \caption{}
      \label{fig:ceal-a}
   \end{subfigure}
   \qquad
   \begin{subfigure}{0.35\textwidth}
      \includegraphics[width=\textwidth]{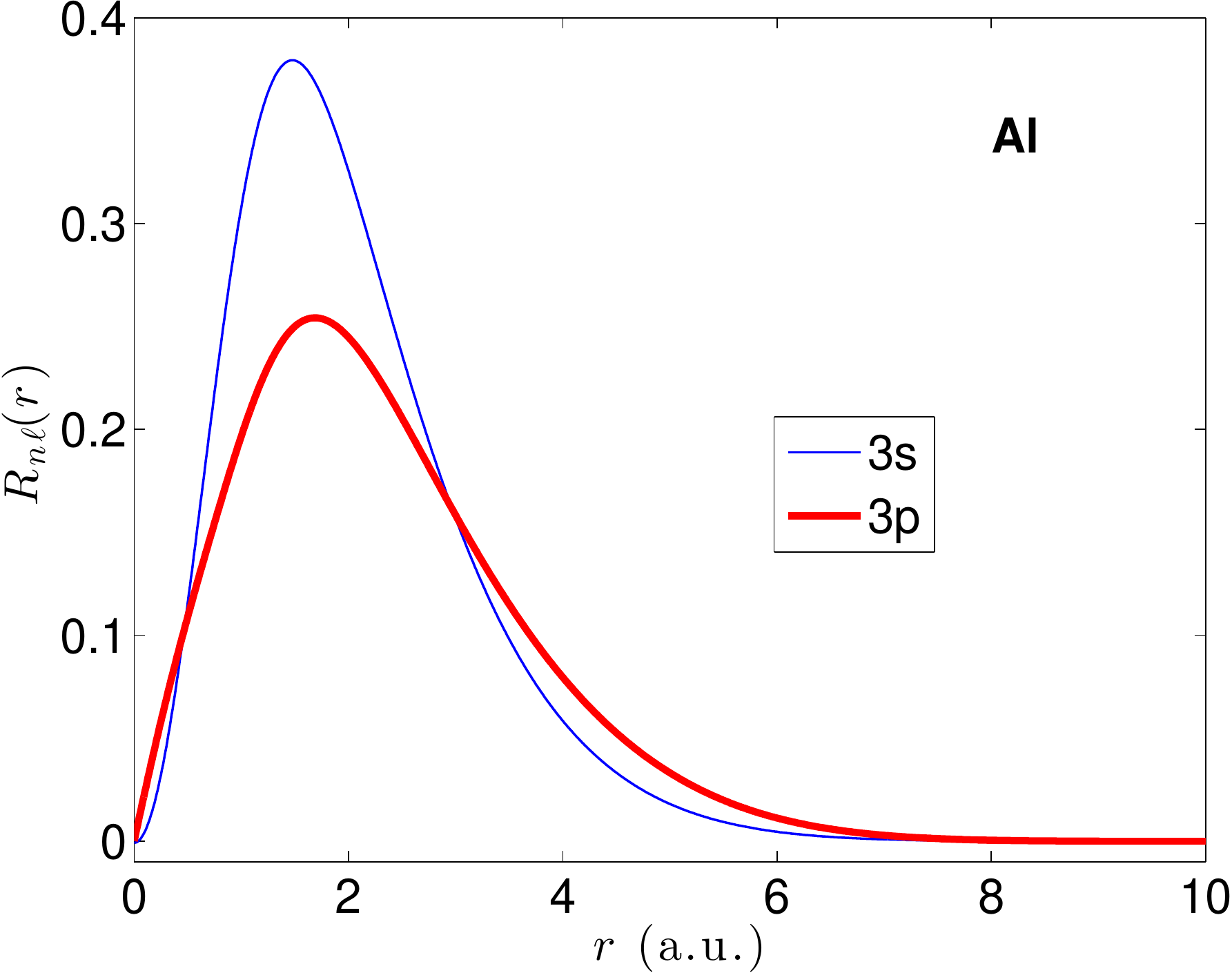}
      \caption{}
      \label{fig:ceal-b}
   \end{subfigure}
   \bigskip
   \begin{subfigure}{0.48\textwidth}
      \includegraphics[width=\textwidth]{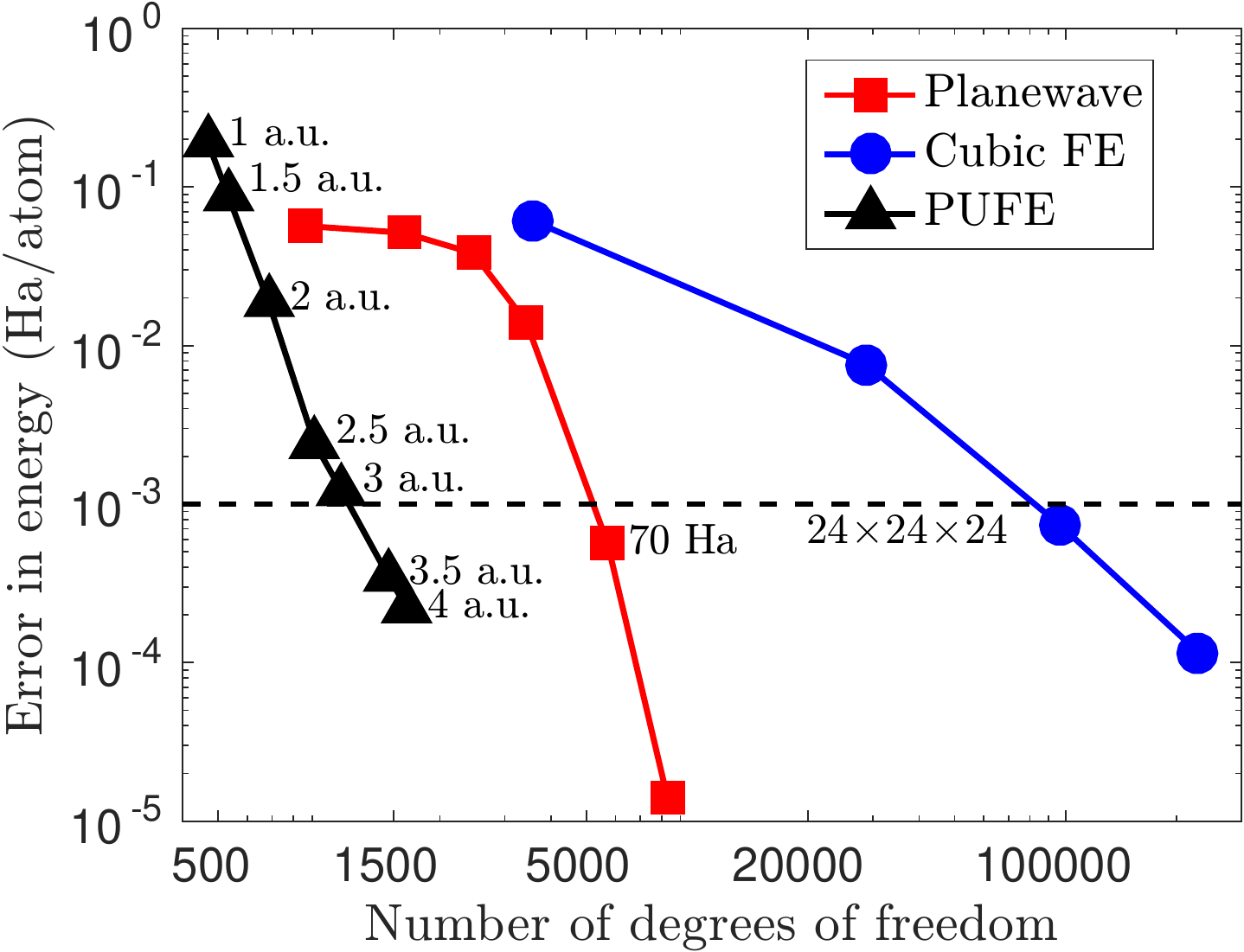}
      \caption{}
      \label{fig:ceal-c}
   \end{subfigure}
\caption{Enrichment functions and convergence of total energy for Kohn-Sham calculations of CeAl. Radial parts of enrichment functions for 
(a) Ce and (b) Al. (c) Convergence of computed total energy per atom for PW, real-space FE, and real-space PUFE methods versus number of degrees of freedom. PUFE calculations use a $4 \times 4 \times 4$ FE mesh. Enrichment support radii are indicated next to corresponding data points for PUFE. Planewave cutoff and FE mesh for $\sim 10^{-3}$ Ha/atom error are indicated next to corresponding points.}\label{fig:ceal}
\end{figure}

Because Ce has a full complement of s, p, d, and f states in valence,
it requires 17 enrichment functions to 
span the occupied space (whereas, e.g., Li requires only two), making this a 
particularly severe test for the efficiency of PUFE relative to planewaves.
The cutoff radius for all enrichment functions was 10 a.u. 
The radial parts of the enrichment functions for Ce and Al
are shown in Figs.~\ref{fig:ceal-a} and~\ref{fig:ceal-b},
respectively.  
The Brillouin zone is sampled at the same two $\kv$-points used for LiH.
The reference result for the total energy is  $-40.675698$ Ha,
taken from a planewave calculation with 170 Ha planewave cutoff. 
Figure~\ref{fig:ceal-c} shows the error in the total 
energy for CeAl versus number of degrees of freedom using 
PW, real-space FE, and real-space PUFE methods. 
We first note that PW calculations require a factor of 16 fewer DOFs to reduce the error to the
desired level than the 
real-space cubic FE method.
We now consider the cubic PUFE result. Remarkably, even with the large number of enrichment functions for Ce, the PUFE method still attains the required 
accuracy with a factor of 5 fewer DOFs than
the current state-of-the-art PW method.

\subsubsection{EOS of LiH}
\begin{figure}
\centering
\includegraphics[width=0.48\textwidth]{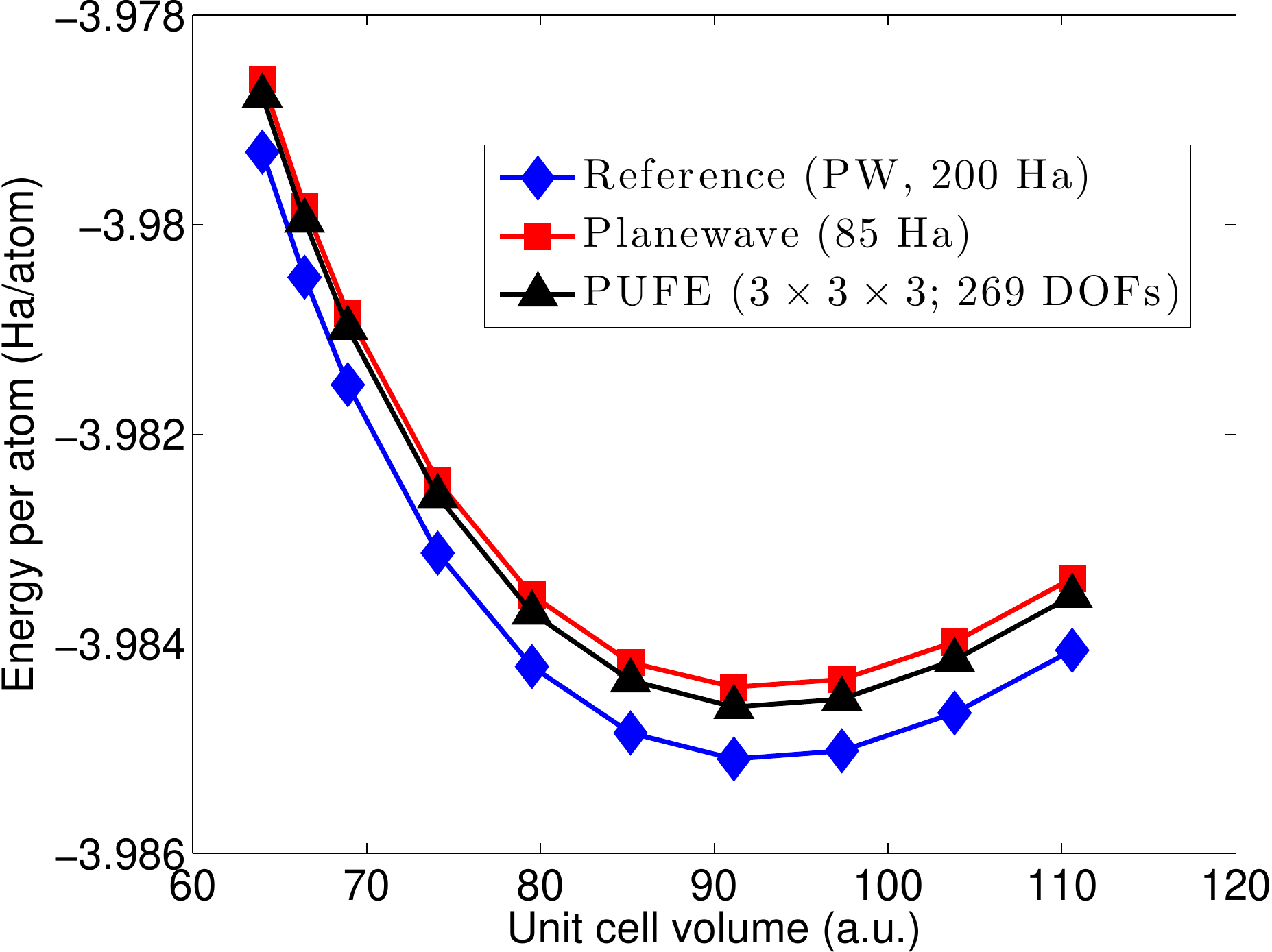}
\caption{EOS of LiH computed by PW and PUFE methods.}\label{fig:EOS-LiH}
\end{figure}

As a last example, 
we compute the equation of state
(EOS) of LiH. The model of the unit cell for LiH and 
the enrichment functions for Li and H are shown in~\fref{fig:LiH}. 
The EOSs computed by PUFE, PW at 85 Ha cutoff, and PW at 200 Ha cutoff are  
shown in~\fref{fig:EOS-LiH}. The error for the PW calculation at
200 Ha cutoff is $\sim 10^{-6}$ Ha/atom, and so is taken as the reference.
The PW calculation at 85 Ha cutoff required from 2398 basis functions
at the smallest volume to 4132 basis functions at the largest volume
to attain the required accuracy. In contrast, the PUFE calculation
required just 269 basis functions to attain the required accuracy at
\textit{all} volumes, a more than order of magnitude reduction in
total DOFs. The computed lattice constants and bulk moduli, 
obtained from a 4th order Birch-Murnaghan fit~\cite{Bir47} are presented
in~\tref{table:EOS}. Both lattice constant and bulk modulus show
excellent agreement with the PW benchmark, while obtained with a basis $\sim$ 1/40 the size.

\begin{table}[H]
\caption{Equilibrium volume, lattice parameter, and bulk modulus of 
         LiH computed by PW and PUFE methods.}\label{table:EOS}
\centering
\begin{tabular}{c c c c} 
\hline
\hline
Method & $V_0$ (Bohr$^3$) & $a_0$ (Bohr) & $B_0$ (GPa) \\ 
\hline
PW (200 Ha)   & 92.5719 & 4.5237 & 45.6656 \\ 
PW (85 Ha)    & 92.5885 & 4.5240 & 45.8108 \\
PUFE          & 92.6241 & 4.5245 & 46.8282 \\ 
\hline
\hline
\end{tabular}
\end{table}

\section{Concluding remarks}\label{sec:conclusions}
In this paper, we have shown that a strictly local, 
systematically improvable, real-space basis can attain the accuracies 
required in quantum mechanical materials calculations with not only fewer 
but substantially fewer degrees of freedom than current 
state-of-the-art planewave based methods.
This was achieved by building known atomic physics into the solution 
process using modern partition-of-unity techniques in finite element analysis. 
Our results show order-of-magnitude reductions in 
basis size relative to state-of-the-art PWs for a range of problems, 
especially those involving localized states. The method developed herein is 
completely general, applicable to any crystal symmetry and to both metals and 
insulators alike. The accuracy and rate of convergence were assessed
on simple systems with light atoms (LiH) as well as a complex f-electron
system (CeAl), which required a large number of atomic-orbital enrichments.
In both cases, the PUFE method attained the required accuracies 
with 5 to 10 times fewer degrees of freedom
than planewave-based methods.  We computed the EOS of LiH
and extracted its lattice constant and bulk modulus, finding again
excellent agreement with benchmark planewave results, while requiring 
an order of magnitude fewer DOFs to obtain.

Having overcome the substantial degree-of-freedom disadvantage of real-space methods relative to current state-or-the-art PW methods, while retaining both strict locality and systematic improvability, there remain two key issues to be addressed: computational cost per degree of freedom and parallel implementation.
First, to be competitive with existing parallel planewave codes in terms of time to solution will require an efficient parallel implementation which exploits the strict locality of the PUFE basis in real space and associated freedom from communication-intensive transforms such as FFTs.
Second, to reduce computational cost per degree of freedom will require efficient solution of the sparse generalized eigenproblem generated by the method, which can become ill-conditioned.
An algorithm for ill-conditioned generalized eigenproblems has
been developed by Cai and coworkers \cite{Cai:2013:HPI} and has shown excellent efficiency
when applied to PUFE system matrices but requires a sparse-direct factorization in its present formulation, limiting problem size.
However, possibilities exist to avoid the ill-conditioned generalized eigenproblem as well.
One such possibility is to employ 
{\em flat-top} partition-of-unity 
functions~\cite{Schweitzer:2011:SEL,Schweitzer:2013:VML}, 
which can be constructed to mitigate ill-conditioning and deliver a
standard, rather than generalized,  eigenproblem. 
Another possibility is to move to a discontinuous representation~\cite{Lin:2012:ALB}, thus 
admitting any desired enrichment while retaining both orthonormality and strict locality. 
Both directions are being pursued by the authors presently.

\section{Acknowledgments}
This work was performed, in part, under the auspices of the U.S.~Department of Energy by
Lawrence Livermore National Laboratory under Contract DE-AC52-07NA27344. 
Support of the Laboratory Directed Research and Development program at the 
Lawrence Livermore National Laboratory is gratefully acknowledged. 
Additional financial support of the National Science Foundation through 
contract grant DMS-0811025 and the
UC Lab Fees Research Program are also acknowledged. 
The authors are also grateful for the initial 
contributions of Murat Efe Guney, Wei Hu, and Seyed
Ebrahim Mousavi in the implementation of the PUFE KS-DFT code.

\appendix
\begin{center}
\section{Supplementary Materials}
\end{center}

\subsection{Partition-of-unity finite-element method applied to a 1D model problem}

\begin{figure}
\centering
\includegraphics[width=0.48\textwidth]{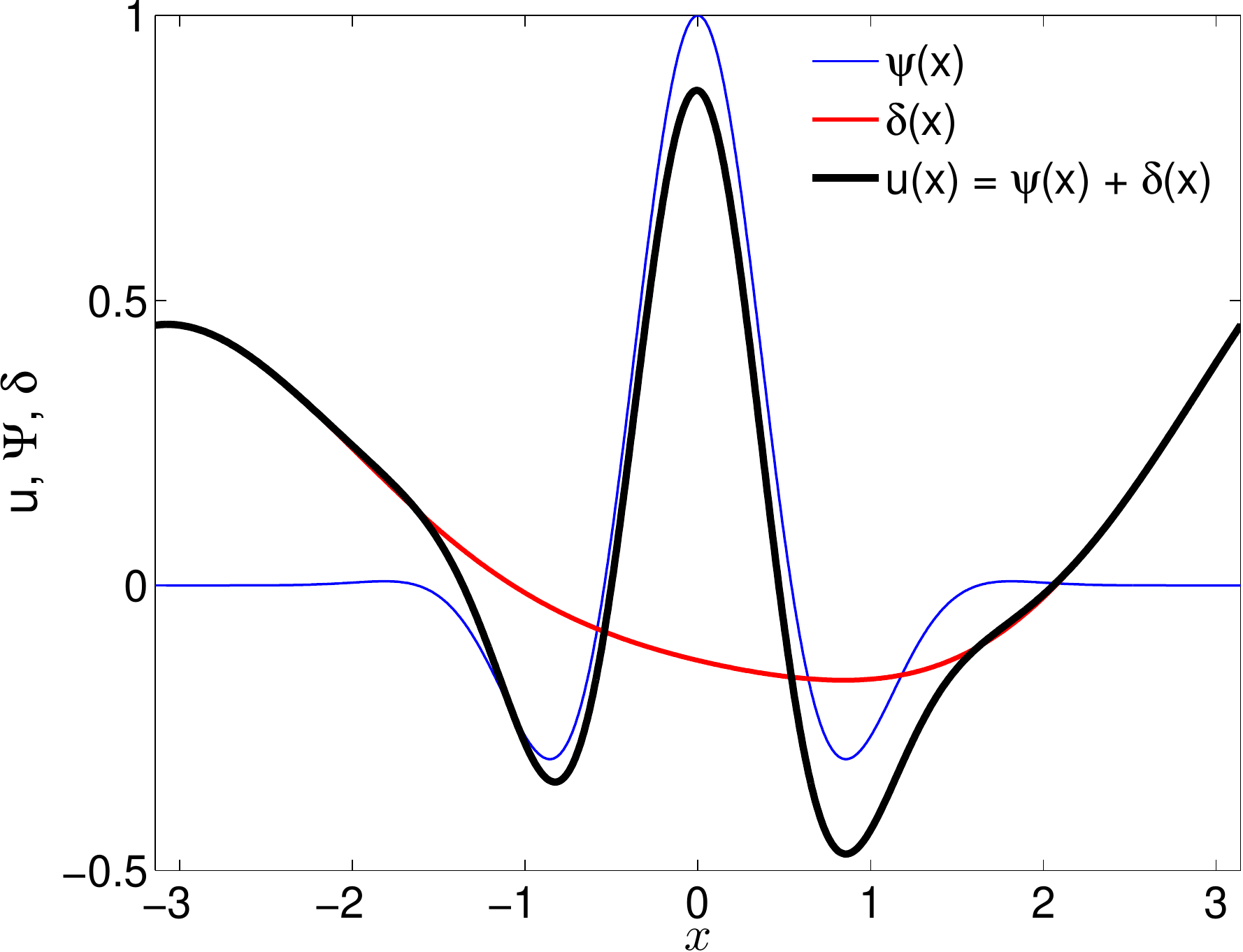}
\caption{Model wavefunction $u(x) = \Psi(x) + \delta(x)$ in a solid, composed of 
isolated-atom like wavefunction $\Psi(x)$ and perturbation $\delta(x)$ 
due to interactions with neighboring atoms.}\label{fig:uexact}
\end{figure}

\begin{figure}[H]
\centering
  \begin{subfigure}{0.36\textwidth}
     \includegraphics[width=\textwidth]{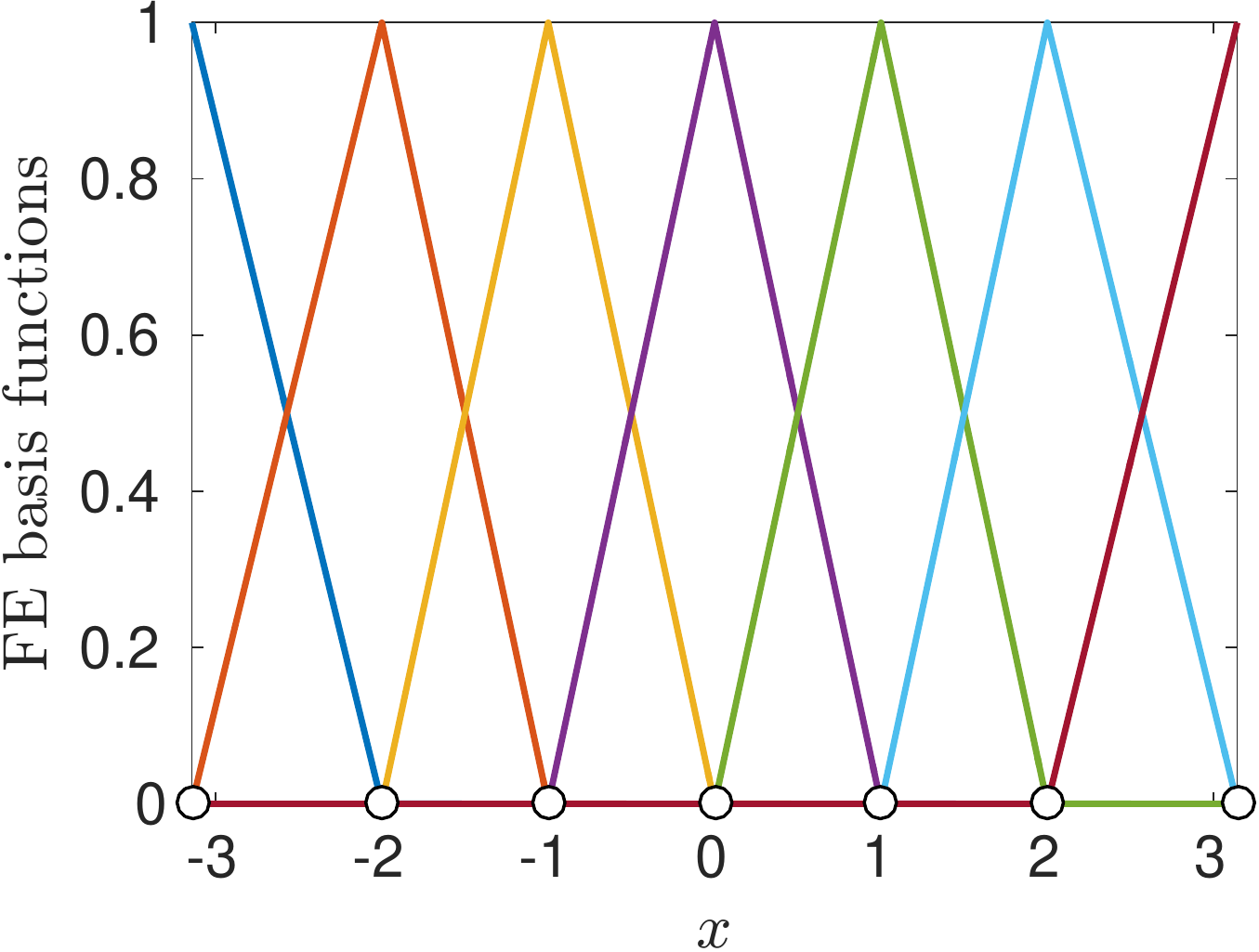}
     \caption{} 
     \label{fig:dataapprox-a}
  \end{subfigure}
  \qquad
  \begin{subfigure}{0.36\textwidth}
     \includegraphics[width=\textwidth]{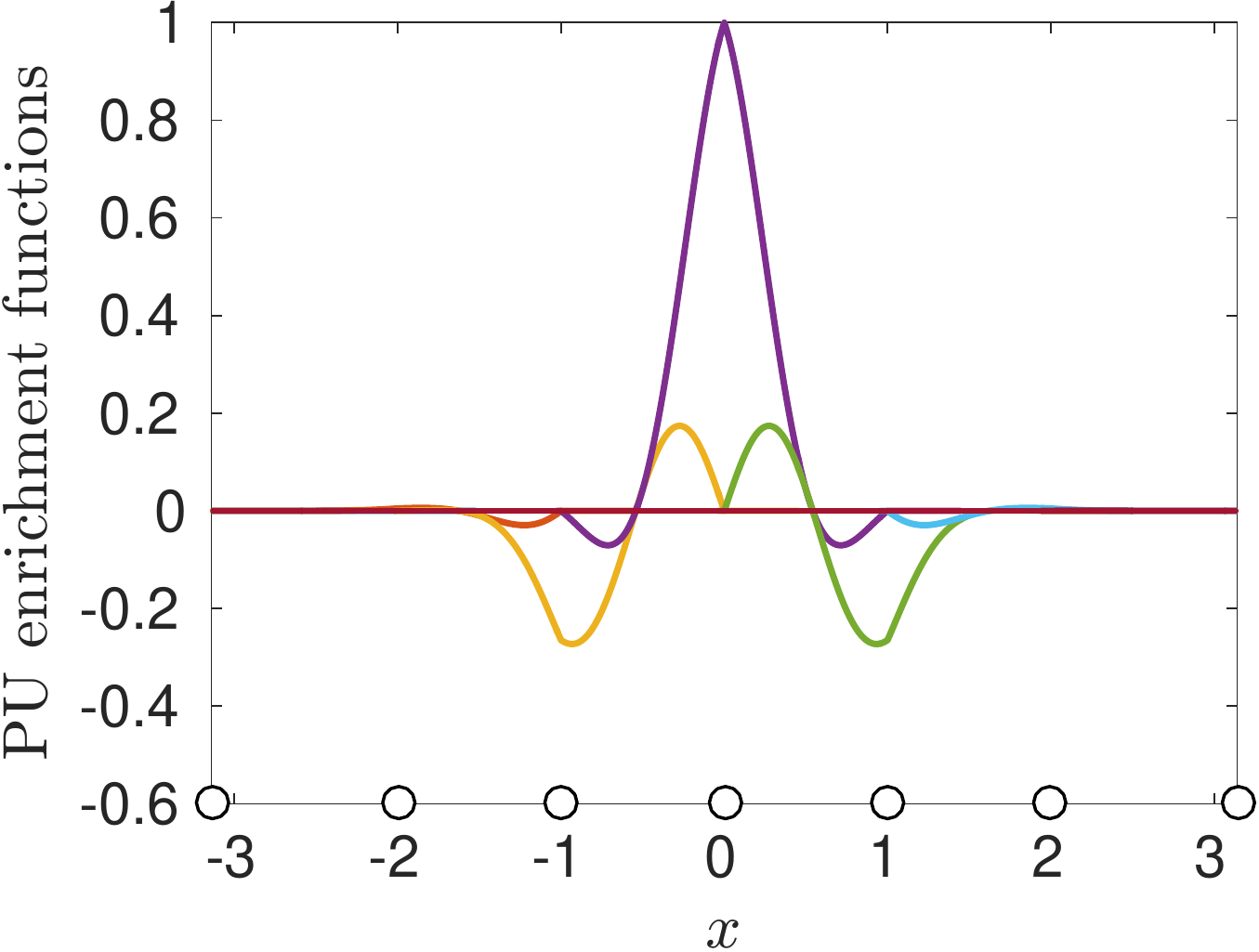}
     \caption{} 
     \label{fig:dataapprox-b}
  \end{subfigure}
  \begin{subfigure}{0.36\textwidth}
     \includegraphics[width=\textwidth]{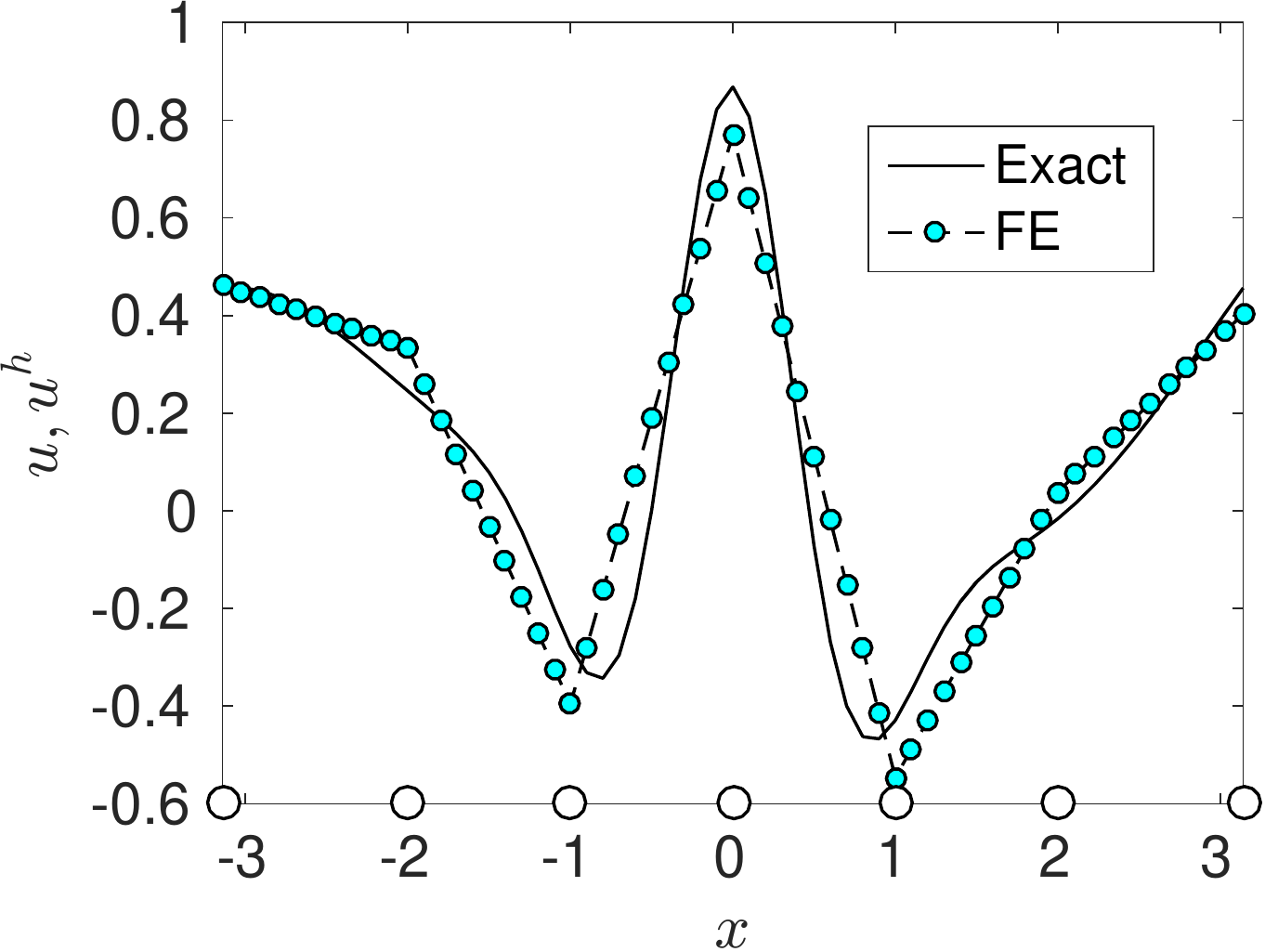}
     \caption{} 
     \label{fig:dataapprox-c}
  \end{subfigure}
  \qquad
  \begin{subfigure}{0.36\textwidth}
     \includegraphics[width=\textwidth]{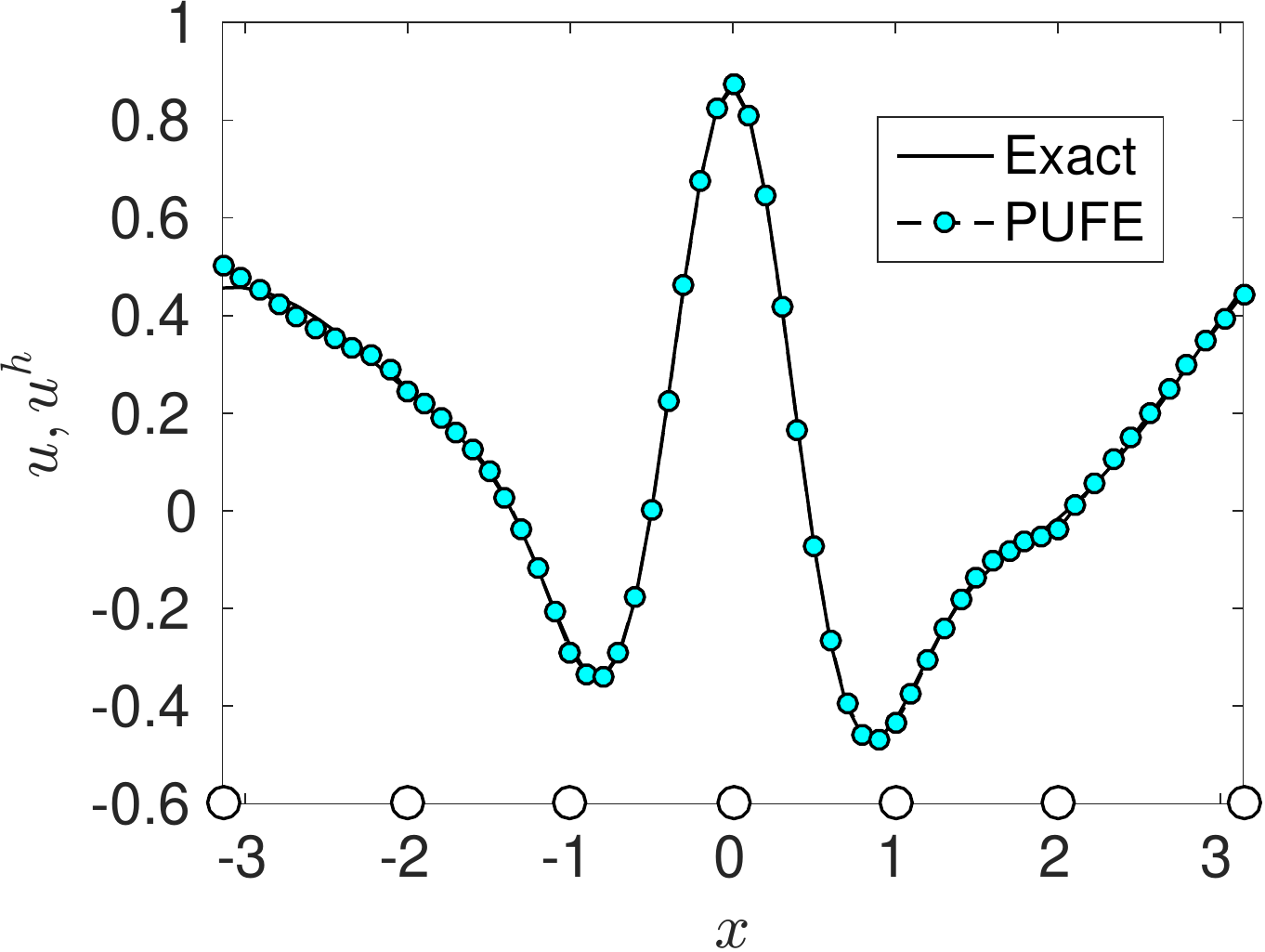}
     \caption{} 
     \label{fig:dataapprox-d}
  \end{subfigure}
  \caption{FE and PUFE approximations of model wavefunction. 
           (a) FE basis functions. (b) PU enrichment functions.
           (c) FE approximation. (d) PUFE approximation.}
   \label{fig:dataapprox}
\end{figure}

To illustrate the substantial advantages of partition-of-unity enrichment 
in the context of problems where much is known about solutions locally, 
as in quantum mechanical calculations of molecular and condensed matter systems, 
we consider a 1D model problem.
Let the one-dimensional domain $\Omega = [-\pi,\pi]$.
We consider the approximation of 
a one-dimensional function $u(x) : \Omega \rightarrow \Re$ representative of 
a quantum-mechanical wavefunction in a molecule or solid (see~\fref{fig:uexact}):
\begin{subequations}
\begin{align}
\label{eq:u}
u(x) &= \Psi(x) + \delta(x), \quad x \in [-\pi,\pi] \\
\intertext{where }
\label{eq:Psi1d}
\Psi(x) &= \exp(-ax^2) \cos (bx), \quad x \in  [-\pi,\pi],\\
\label{eq:delta}
\delta(x) &= c(1+dx^2)\cos(x - 0.5), \quad x \in  [-\pi,\pi],
\end{align}
\end{subequations}
and we choose $a = 1.3$, $b = 2.9$, $c = -0.15$, $d=0.25$. 
In~\eref{eq:delta}, 
$\delta(x)$ represents a perturbation arising from interactions with neighboring atoms. Let the
domain be discretized by six elements as shown in~\fref{fig:dataapprox-a}. 
The coordinates of the seven nodes are:
$x_1 = -\pi$, $x_2 = -2$, $x_3 = -1$, $x_4=0$,
$x_5 = 1$, $x_6 = 2$, $x_7 = \pi$. Now, consider the following
PUFE approximation:
\begin{equation*}
u^h (x) = \sum_{i=1}^7 \phi_i(x) u_i +
\sum_{i=1}^7 \phi_i(x) \Psi(x) b_i
\equiv \sum_{j=1}^{14} \varphi_j(x) c_j ,
\end{equation*}
where $\phi_i(x)$ are linear finite element basis functions, 
$\Psi(x)$ is the enrichment function given in \eref{eq:Psi1d}, 
and the functions $\phi_i(x)$ serve as both standard FE basis and partition of unity 
for enrichment function $\Psi(x)$.
If $b_i = 0$ for all $i$, then we recover the standard FE approximation.
To compute these coefficients for both FE and PUFE approximations, 
we construct the $L^2$ projection of $u(x)$ onto the space
spanned by the PUFE basis $\{\varphi_j(x)\}$.
Let
\begin{equation*}
(u,w) \equiv (u,w)_2 := \int_{-\pi}^\pi uw \, dx
\end{equation*}
denote the $L^2$ inner product for functions $u,w \in L^2(-\pi,\pi)$.
Then, the best-fit to $u(x)$ is obtained by minimizing 
$(u-u^h,u-u^h)/2$.  Since
\begin{equation*} 
\tfrac{1}{2} (u - u^h,u - u^h) 
= \tfrac{1}{2} (u,u) - (u,u^h) + \tfrac{1}{2}(u^h,u^h), 
\end{equation*} 
we can equivalently minimize $\Pi(u,u^h) = (u^h,u^h)/2 - (u,u^h)$. Now,
\begin{align*}
\tfrac{1}{2}(u^h,u^h) &= \tfrac{1}{2} \vm{c}^T \vm{M} \vm{c}, \quad
\vm{M}_{ij} = \int_{-\pi}^\pi \varphi_i(x) \varphi_j(x) \, dx, \\
(u,u^h) &= \vm{c}^T \vm{f}, \quad
\vm{f}_i = \int_{-\pi}^\pi \varphi_i(x) u(x) \, dx, \quad
\vm{c} = [ c_1 \dotsc c_{14} ]^T.
\end{align*}
Therefore, setting $\partial \Pi / \partial \vm{c} = \vm{0}$ leads to
the following linear system of equations:
\begin{equation}
\vm{M}\vm{c} = \vm{f},
\end{equation}
which is solved to obtain the minimizing coefficients $\vm{c}$. 

The bases and resulting approximations are shown in
\fref{fig:dataapprox}. The classical FE basis functions $\phi_i(x)$ and 
PU enrichment functions $\phi_i(x)\Psi(x)$ are shown in 
Figs.~\ref{fig:dataapprox-a} and \ref{fig:dataapprox-b}, respectively. 
The PUFE basis then consists of the classical basis functions $\phi_i(x)$ 
and PU enrichment functions $\phi_i(x)\Psi(x)$ together. 
On comparing the FE and PUFE approximations shown in
Figs.~\ref{fig:dataapprox-c} and~\ref{fig:dataapprox-d}, 
the significant advantage of the PUFE basis is clear: 
the accuracy of the standard FE approximation is dramatically improved by
adding the PU enrichment functions to the basis.

\bibliographystyle{unsrtnat}
\bibliography{pufe-dft.bib}

\begin{thebibliography}{75}
\providecommand{\natexlab}[1]{#1}
\providecommand{\url}[1]{\texttt{#1}}
\expandafter\ifx\csname urlstyle\endcsname\relax
  \providecommand{\doi}[1]{doi: #1}\else
  \providecommand{\doi}{doi: \begingroup \urlstyle{rm}\Url}\fi

\bibitem[Hohenberg and Kohn(1964)]{HohK64}
P.~Hohenberg and W.~Kohn.
\newblock Inhomogeneous electron gas.
\newblock \emph{Phys. Rev.}, 136:\penalty0 B864--871, 1964.

\bibitem[Kohn and Sham(1965)]{KohS65}
W.~Kohn and L.~J. Sham.
\newblock Self-consistent equations including exchange and correlation effects.
\newblock \emph{Phys. Rev.}, 140:\penalty0 A1133--1138, 1965.

\bibitem[Pickett(1989)]{Pic89}
W.~E. Pickett.
\newblock Pseudopotential methods in condensed matter applications.
\newblock \emph{Computer Phys.\ Rep.}, 9\penalty0 (3):\penalty0 115--197, 1989.

\bibitem[Arias(1999)]{Ar99}
T.~A. Arias.
\newblock Multiresolution analysis of electronic structure: Semicardinal and
  wavelet bases.
\newblock \emph{Rev. Mod. Phys.}, 71\penalty0 (1):\penalty0 267--311, 1999.

\bibitem[Beck(2000)]{Bec00}
T.~L. Beck.
\newblock Real-space mesh techniques in density-functional theory.
\newblock \emph{Rev. Mod. Phys.}, 72\penalty0 (4):\penalty0 1041--1080, 2000.

\bibitem[Torsti et~al.(2006)Torsti, Eirola, Enkovaara, Hakala, Havu, Havu,
  H\"oyn\"al\"anmaa, Ignatius, Lyly, Makkonen, Rantala, Ruokolainen,
  Ruotsalainen, R\"as\"anen, Saarikoski, and Puska]{TorEE06}
T.~Torsti, T.~Eirola, J.~Enkovaara, T.~Hakala, P.~Havu, V.~Havu,
  T.~H\"oyn\"al\"anmaa, J.~Ignatius, M.~Lyly, I.~Makkonen, T.~T. Rantala,
  J.~Ruokolainen, K.~Ruotsalainen, E.~R\"as\"anen, H.~Saarikoski, and M.~J.
  Puska.
\newblock Three real-space discretization techniques in electronic structure
  calculations.
\newblock \emph{Phys. Stat. Sol. (B)}, 243\penalty0 (5):\penalty0 1016--1053,
  2006.

\bibitem[Pask and Sterne(2005{\natexlab{a}})]{PasS05b}
J.~E. Pask and P.~A. Sterne.
\newblock Finite element methods in ab initio electronic structure
  calculations.
\newblock \emph{Model. Simul. Mater. Sci. Eng.}, 13\penalty0 (3):\penalty0
  R71--R96, 2005{\natexlab{a}}.

\bibitem[Genovese et~al.({2008})Genovese, Neelov, Goedecker, Deutsch, Ghasemi,
  Willand, Caliste, Zilberberg, Rayson, Bergman, and Schneider]{GenNG08}
L.~Genovese, A.~Neelov, S.~Goedecker, T.~Deutsch, S.~A. Ghasemi, A.~Willand,
  D.~Caliste, O.~Zilberberg, M.~Rayson, A.~Bergman, and R.~Schneider.
\newblock {Daubechies wavelets as a basis set for density functional
  pseudopotential calculations}.
\newblock \emph{{J. Chem. Phys.}}, {129}\penalty0 ({1}):\penalty0 {014109},
  {2008}.

\bibitem[Saad et~al.(2010)Saad, Chelikowsky, and Shontz]{Saad:2010:NMF}
Y.~Saad, J.~R. Chelikowsky, and S.~M. Shontz.
\newblock Numerical methods for electronic structure calculations of materials.
\newblock \emph{SIAM Rev.}, 52\penalty0 (1):\penalty0 3--54, 2010.

\bibitem[Chelikowsky et~al.(1994)Chelikowsky, Troullier, and Saad]{ChelTS94}
J.~R. Chelikowsky, N.~Troullier, and Y.~Saad.
\newblock Finite-difference-pseudopotential method: Electronic-structure
  calculations without a basis.
\newblock \emph{Phys. Rev. Lett.}, 72\penalty0 (8):\penalty0 1240--1243, 1994.

\bibitem[Seitsonen et~al.(1995)Seitsonen, Puska, and Nieminen]{SeitPN95}
A.~P. Seitsonen, M.~J. Puska, and R.~M. Nieminen.
\newblock Real-space electronic-structure calculations: Combination of the
  finite-difference and conjugate-gradient methods.
\newblock \emph{Phys. Rev. B}, 51\penalty0 (20):\penalty0 14057--14061, 1995.

\bibitem[Gygi and Galli(1995)]{GygG95}
F.~Gygi and G.~Galli.
\newblock Real-space adaptive-coordinate electronic-structure calculations.
\newblock \emph{Phys. Rev. B}, 52\penalty0 (4):\penalty0 R2229--R2232, 1995.

\bibitem[Iyer et~al.(1995)Iyer, Merrick, and Beck]{IyMB95}
K.~A. Iyer, M.~P. Merrick, and T.~L. Beck.
\newblock Application of a distributed nucleus approximation in grid based
  minimization of the {K}ohn-{S}ham energy functional.
\newblock \emph{J. Chem. Phys.}, 103\penalty0 (1):\penalty0 227--233, 1995.

\bibitem[Hoshi et~al.(1995)Hoshi, Arai, and Fujiwara]{HosAF95}
T.~Hoshi, M.~Arai, and T.~Fujiwara.
\newblock Density-functional molecular-dynamics with real-space
  finite-difference.
\newblock \emph{Phys. Rev. B}, 52\penalty0 (8):\penalty0 R5459--R5462, 1995.

\bibitem[Briggs et~al.(1996)Briggs, Sullivan, and Bernholc]{BrigSB96}
E.~L. Briggs, D.~J. Sullivan, and J.~Bernholc.
\newblock Real-space multigrid-based approach to large-scale electronic
  structure calculations.
\newblock \emph{Phys. Rev. B}, 54\penalty0 (20):\penalty0 14362--14375, 1996.

\bibitem[Modine et~al.(1997)Modine, Zumbach, and Kaxiras]{ModZK97}
N.~A. Modine, G.~Zumbach, and E.~Kaxiras.
\newblock Adaptive-coordinate real-space electronic-structure calculations for
  atoms, molecules, and solids.
\newblock \emph{Phys. Rev. B}, 55\penalty0 (16):\penalty0 10289--10301, 1997.

\bibitem[Fattebert(1999)]{Fat99}
J.-L. Fattebert.
\newblock Finite difference schemes and block {Rayleigh Quotient Iteration} for
  electronic structure calculations on composite grids.
\newblock \emph{J. Comp.\ Phys.}, 149\penalty0 (1):\penalty0 75--94, 1999.

\bibitem[Fattebert and Bernholc(2000)]{FB:1}
J.-L. Fattebert and J.~Bernholc.
\newblock Towards grid-based {O(N)} density-functional theory methods:
  optimized non-orthogonal orbitals and multigrid acceleration.
\newblock \emph{Phys. Rev. B}, 62\penalty0 (3):\penalty0 1713--1722, 2000.

\bibitem[Fattebert and Gygi(2004)]{FG:3}
J.-L. Fattebert and F.~Gygi.
\newblock Linear scaling first-principles molecular dynamics with controlled
  accuracy.
\newblock \emph{Comput. Phys. Comm.}, 162:\penalty0 24--36, 2004.

\bibitem[Alemany et~al.(2004)Alemany, Jain, Kronik, and Chelikowsky]{AlJK04}
M.~M.~G. Alemany, M.~Jain, L.~Kronik, and J.~R. Chelikowsky.
\newblock Real-space pseudopotential method for computing the electronic
  properties of periodic systems.
\newblock \emph{Phys. Rev. B}, 69\penalty0 (7):\penalty0 075101, 2004.

\bibitem[Beck({2009})]{Beck:2009:RSM}
T.~L. Beck.
\newblock {Real-space and multigrid methods in computational chemistry}.
\newblock In K.~B. Lipkowitz and T.~R. Cundari, editors, \emph{{Reviews in
  Computational Chemistry}}, volume~{26}, pages {223--285}. {2009}.

\bibitem[Ghosh and Suryanarayana(2016{\natexlab{a}})]{Ghosh:2016:SPARC1}
S.~Ghosh and P.~Suryanarayana.
\newblock {SPARC}: Accurate and efficient finite-difference formulation and
  parallel implementation of density functional theory: Isolated clusters.
\newblock \emph{arXiv preprint arXiv:1603.04334}, 2016{\natexlab{a}}.

\bibitem[Ghosh and Suryanarayana(2016{\natexlab{b}})]{Ghosh:2016:SPARC2}
S.~Ghosh and P.~Suryanarayana.
\newblock {SPARC}: Accurate and efficient finite-difference formulation and
  parallel implementation of density functional theory: {Extended} systems.
\newblock \emph{arXiv preprint arXiv:1603.04339}, 2016{\natexlab{b}}.

\bibitem[Ono and Hirose(1999)]{On99}
T.~Ono and K.~Hirose.
\newblock Timesaving double-grid method for real-space electronic-structure
  calculations.
\newblock \emph{Phys. Rev. Lett.}, 82\penalty0 (25):\penalty0 5016--5019, 1999.

\bibitem[Strang and Fix(1973)]{StranF73}
G.~Strang and G.~J. Fix.
\newblock \emph{An Analysis of the Finite Element Method}.
\newblock Prentice-Hall, Englewood Cliffs, 1973.

\bibitem[Askar(1975)]{As75}
A.~Askar.
\newblock Finite-element method for bound-state calculations in
  quantum-mechanics.
\newblock \emph{J. Chem. Phys.}, 62\penalty0 (2):\penalty0 732--734, 1975.

\bibitem[White et~al.(1989)White, Wilkins, and Teter]{WhitWT89}
S.~R. White, J.~W. Wilkins, and M.~P. Teter.
\newblock Finite-element method for electronic-structure.
\newblock \emph{Phys. Rev. B}, 39\penalty0 (9):\penalty0 5819--5833, 1989.

\bibitem[Hermansson and Yevick(1986)]{HerY86}
B.~Hermansson and D.~Yevick.
\newblock Finite-element approach to band-structure analysis.
\newblock \emph{Phys. Rev. B}, 33\penalty0 (10):\penalty0 7241--7242, 1986.

\bibitem[Tsuchida and Tsukada(1995)]{TsucT95a}
E.~Tsuchida and M.~Tsukada.
\newblock Electronic-structure calculations based on the finite-element method.
\newblock \emph{Phys. Rev. B}, 52\penalty0 (8):\penalty0 5573--5578, 1995.

\bibitem[Tsuchida and Tsukada(1998)]{TsucT98}
E.~Tsuchida and M.~Tsukada.
\newblock Large-scale electronic-structure calculations based on the adaptive
  finite-element method.
\newblock \emph{J. Phys. Soc. Jpn.}, 67\penalty0 (11):\penalty0 3844--3858,
  1998.

\bibitem[Tsuchida and Tsukada(1996)]{TsucT96}
E.~Tsuchida and M.~Tsukada.
\newblock Adaptive finite-element method for electronic-structure calculations.
\newblock \emph{Phys. Rev. B}, 54\penalty0 (11):\penalty0 7602--7605, 1996.

\bibitem[Pask et~al.(1999)Pask, Klein, Fong, and Sterne]{PasKF99}
J.~E. Pask, B.~M. Klein, C.~Y. Fong, and P.~A. Sterne.
\newblock Real-space local polynomial basis for solid-state
  electronic-structure calculations: A finite-element approach.
\newblock \emph{Phys. Rev. B}, 59\penalty0 (19):\penalty0 12352--12358, 1999.

\bibitem[Pask et~al.(2001)Pask, Klein, Sterne, and Fong]{PasKS01}
J.~E. Pask, B.~M. Klein, P.~A. Sterne, and C.~Y. Fong.
\newblock Finite-element methods in electronic-structure theory.
\newblock \emph{Comput. Phys. Commun.}, 135\penalty0 (1):\penalty0 1--34, 2001.

\bibitem[Pask and Sterne(2005{\natexlab{b}})]{PasS05a}
J.~E. Pask and P.~A. Sterne.
\newblock Real-space formulation of the electrostatic potential and total
  energy of solids.
\newblock \emph{Phys. Rev. B}, 71\penalty0 (11):\penalty0 113101,
  2005{\natexlab{b}}.

\bibitem[Motamarri et~al.(2013)Motamarri, Nowak, Leiter, Knap, and
  Gavini]{Motamarri:2013:HOA}
P.~Motamarri, M.~Nowak, K.~Leiter, J.~Knap, and V.~Gavini.
\newblock Higher-order adaptive finite-element methods for {Kohn-Sham} density
  functional theory.
\newblock \emph{J. Comp.\ Phys.}, 253:\penalty0 308--343, 2013.

\bibitem[Tsuchida et~al.(2015)Tsuchida, Choe, and Ohkubo]{Tsuc15}
E.~Tsuchida, Y.-K. Choe, and T.~Ohkubo.
\newblock Adaptive finite-element method for large-scale {\em ab initio\/}
  molecular dynamics simulations.
\newblock \emph{Phys. Chem. Chem. Phys.}, 17\penalty0 (47):\penalty0
  31444--31452, 2015.

\bibitem[Batcho({2000})]{Bat00}
P.~F. Batcho.
\newblock {Computational method for general multicenter electronic structure
  calculations}.
\newblock \emph{{Phys. Rev. E}}, {61}\penalty0 ({6, Part B}):\penalty0
  {7169--7183}, {2000}.

\bibitem[Lehtovaara et~al.({2009})Lehtovaara, Havu, and Puska]{LehHP09}
L.~Lehtovaara, V.~Havu, and M.~Puska.
\newblock {All-electron density functional theory and time-dependent density
  functional theory with high-order finite elements}.
\newblock \emph{{J. Chem. Phys.}}, {131}\penalty0 ({5}):\penalty0 {054103},
  {2009}.

\bibitem[Zhang et~al.(2008)Zhang, Shen, Zhou, and Gong]{Zhang:2008:FEM}
D.~Zhang, L.~H. Shen, A.~H. Zhou, and X.~G. Gong.
\newblock Finite element method for solving {Kohn-Sham} equations based on
  self-adaptive tetrahedral mesh.
\newblock \emph{Phys.\ Lett.\ A}, 372:\penalty0 5071--5076, 2008.

\bibitem[Bylaska et~al.(2009)Bylaska, Holst, and Weare]{BylHW09}
E.~J. Bylaska, M.~Holst, and J.~H. Weare.
\newblock Adaptive finite element method for solving the exact {Kohn-Sham}
  equation of density functional theory.
\newblock \emph{J. Chem. Theory Comput.}, 5\penalty0 (4):\penalty0 937--948,
  2009.

\bibitem[Alizadegan et~al.(2010)Alizadegan, Hsia, and
  Mart{\'i}nez]{Alizadegan:2010:ADA}
R.~Alizadegan, K.~J. Hsia, and T.~J. Mart{\'i}nez.
\newblock A divide and conquer real space finite-element {Hartree-Fock} method.
\newblock \emph{J. Chem. Phys.}, 132:\penalty0 034101, 2010.

\bibitem[Bao et~al.(2012)Bao, Hu, and Liu]{Bao:2012:HAF}
G.~Bao, G.~Hu, and D.~Liu.
\newblock An $h$-adaptive finite element solver for the calculations of the
  electronic structures.
\newblock \emph{J. Comp. Phys.}, 231\penalty0 (14):\penalty0 4967--4979, 2012.

\bibitem[Schauer and Linder(2013)]{Schauer:2013:AEK}
V.~Schauer and C.~Linder.
\newblock All-electron {Kohn-Sham} density functional theory on hierarchic
  finite element spaces.
\newblock \emph{J. Comp. Phys.}, 250:\penalty0 644--664, 2013.

\bibitem[Schauer and Linder(2015)]{Schauer:2014:RBM}
V.~Schauer and C.~Linder.
\newblock The reduced basis method in all-electron calculations with finite
  elements.
\newblock \emph{Adv.\ Comput.\ Math.}, 41\penalty0 (5):\penalty0 1035--1047,
  2015.

\bibitem[Motamarri and Gavini(2014)]{Motamarri:2014:SSS}
P.~Motamarri and V.~Gavini.
\newblock A subquadratic-scaling subspace projection method for large-scale
  {Kohn-Sham DFT} calculations using spectral finite-element discretization.
\newblock \emph{Phys. Rev.\ B}, 90:\penalty0 115127, 2014.

\bibitem[Chen et~al.(2014)Chen, Dai, Gong, He, and Zhou]{Chen:2014:AFE}
H.~Chen, X.~Dai, X.~Gong, L.~He, and A.~Zhou.
\newblock Adaptive finite element approximations for {Kohn-Sham} models.
\newblock \emph{Multiscale Model.\ Sim.}, 12\penalty0 (4):\penalty0 1828--1869,
  2014.

\bibitem[Maday(2014)]{Maday:2014:HPF}
Y.~Maday.
\newblock {$h$ -- $P$} finite element approximation for full-potential
  electronic structure calculations.
\newblock \emph{Chinese Ann.\ Math.\ B}, 35\penalty0 (1):\penalty0 1--24, 2014.

\bibitem[Davydov et~al.(2016)Davydov, Young, and Steinmann]{Davydov:2016:AFE}
D.~Davydov, T.~D. Young, and P.~Steinmann.
\newblock On the adaptive finite element analysis of the {Kohn-Sham} equations:
  methods, algorithms, and implementation.
\newblock \emph{Internat. J. Numer.\ Methods Engrg.}, 106:\penalty0 863--888,
  2016.

\bibitem[Skriver(1984)]{Skriv84}
H.~L. Skriver.
\newblock \emph{The LMTO Method}.
\newblock Springer, Berlin, 1984.

\bibitem[Singh and Nordstrom(2006)]{SinN06}
D.~J. Singh and L.~Nordstrom.
\newblock \emph{Planewaves, Pseudopotentials, and the LAPW Method}.
\newblock Springer, New York, 2nd edition, 2006.

\bibitem[Boys(1950)]{Boys:1950:EWF}
S.~F. Boys.
\newblock Electronic wave functions. {I}. {A} general method of calculation for
  the stationary states of any molecular system.
\newblock \emph{Proc. R. Soc. Lon. A}, 200\penalty0 (1063):\penalty0 542--554,
  1950.

\bibitem[Dusterhoft et~al.(1998)Dusterhoft, Heinemann, and Kolb]{DusHK98}
C.~Dusterhoft, D.~Heinemann, and D.~Kolb.
\newblock {D}irac-{F}ock-{S}later calculations for diatomic molecules with a
  finite element defect correction method {(FEM-DKM)}.
\newblock \emph{Chem. Phys. Lett.}, 296\penalty0 (1-2):\penalty0 77--83, 1998.

\bibitem[Yamakawa and Hyodo(2003)]{YamH03}
S.~Yamakawa and S.~Hyodo.
\newblock Electronic state calculation of hydrogen in metal clusters based on
  {Gaussian-FEM} mixed basis function.
\newblock \emph{J. Alloy. Compd.}, 356\penalty0 (2):\penalty0 231--235, 2003.

\bibitem[Yamakawa and Hyodo(2005)]{YamH05}
S.~Yamakawa and S.~Hyodo.
\newblock Gaussian finite-element mixed-basis method for electronic structure
  calculations.
\newblock \emph{Phys. Rev. B}, 71\penalty0 (3):\penalty0 035113, 2005.

\bibitem[Jun(2004)]{Jun04}
S.~Jun.
\newblock Meshfree implementation for the real-space electronic-structure
  calculation of crystalline solids.
\newblock \emph{Internat. J. Numer.\ Methods Engrg.}, 59\penalty0
  (14):\penalty0 1909--1923, 2004.

\bibitem[Chen et~al.(2007)Chen, Wu, and Puso]{Che08}
J.~S. Chen, H.~Wu, and M.~A. Puso.
\newblock Orbital {HP}-cloud for solving {S}chr{\"o}dinger equation in quantum
  mechanics.
\newblock \emph{Comput.\ Methods Appl.\ Mech.\ Engrg.}, 196\penalty0
  (37--40):\penalty0 3693--3705, 2007.

\bibitem[Suryanarayana et~al.(2011)Suryanarayana, Bhattacharya, and
  Ortiz]{Surya:2011:AMC}
P.~Suryanarayana, K.~Bhattacharya, and M.~Ortiz.
\newblock A mesh-free convex approximation scheme for {Kohn-Sham} density
  functional theory.
\newblock \emph{J. Comp. Phys.}, 230:\penalty0 5226--5238, 2011.

\bibitem[Sukumar and Pask(2009)]{Sukumar:2009:CEF}
N.~Sukumar and J.~E. Pask.
\newblock Classical and enriched finite element formulations for
  {Bloch}-periodic boundary conditions.
\newblock \emph{Internat. J. Numer.\ Methods Engrg.}, 77\penalty0 (8):\penalty0
  1121--1138, 2009.

\bibitem[Pask et~al.(2012)Pask, Sukumar, and Mousavi]{Pask:2012:LSS}
J.~E. Pask, N.~Sukumar, and S.~E. Mousavi.
\newblock Linear scaling solution of the all-electron {Coulomb} problem in
  solids.
\newblock \emph{Int.\ J. Multiscale Comp.\ Eng.}, 10\penalty0 (1):\penalty0
  83--99, 2012.

\bibitem[Pask et~al.(2011)Pask, Sukumar, Guney, and Hu]{Pask:2011:PUF}
J.~E. Pask, N.~Sukumar, M.~Guney, and W.~Hu.
\newblock Partition-of-unity finite-element method for large scale quantum
  molecular dynamics on massively parallel computational platforms.
\newblock Technical Report LLNL-TR-470692, Department of Energy LDRD Grant
  08-ERD-052, March 2011.

\bibitem[Melenk and Babu\v{s}ka(1996)]{Melenk:1996:PUF}
J.~M. Melenk and I.~Babu\v{s}ka.
\newblock The partition of unity finite element method: Basic theory and
  applications.
\newblock \emph{Comput.\ Methods Appl.\ Mech.\ Engrg.}, 139:\penalty0 289--314,
  1996.

\bibitem[Babu\v{s}ka and Melenk(1997)]{Babuska:1997:PUM}
I.~Babu\v{s}ka and J.~M. Melenk.
\newblock The partition of unity method.
\newblock \emph{Internat. J. Numer.\ Methods Engrg.}, 40:\penalty0 727--758,
  1997.

\bibitem[Perdew et~al.(1996)Perdew, Burke, and Ernzerhof]{perdew:1996:GGA}
J.~P. Perdew, K.~Burke, and M.~Ernzerhof.
\newblock Generalized gradient approximation made simple.
\newblock \emph{Phys.\ Rev.\ Lett.}, 18:\penalty0 3865, 1996.

\bibitem[Martin(2004)]{Mar04}
R.~M. Martin.
\newblock \emph{Electronic Structure: Basic Theory and Practical Methods}.
\newblock Cambridge University Press, Cambridge, 2004.

\bibitem[Ashcroft and Mermin(1976)]{ashcroft:book}
N.~W. Ashcroft and N.~D. Mermin.
\newblock \emph{Solid State Physics}.
\newblock Holt, Rinehart and Winston, New York, 1976.

\bibitem[Griffiths(1994)]{Griffiths:book}
D.~J. Griffiths.
\newblock \emph{Introduction to Quantum Mechanics}.
\newblock Prentice Hall, Englewood Cliffs, 1994.

\bibitem[Mousavi et~al.(2012)Mousavi, Pask, and Sukumar]{Mousavi:2012:EAI}
S.~E. Mousavi, J.~E. Pask, and N.~Sukumar.
\newblock Efficient adaptive integration of functions with sharp gradients and
  cusps in $n$-dimensional parallelepipeds.
\newblock \emph{Internat. J. Numer.\ Methods Engrg.}, 91\penalty0 (4):\penalty0
  343--357, 2012.

\bibitem[Gygi(1992)]{Gyg92}
F.~Gygi.
\newblock Adaptive {R}iemannian metric for plane-wave electronic-structure
  calculations.
\newblock \emph{Europhys. Lett.}, 19\penalty0 (7):\penalty0 617--622, 1992.

\bibitem[Hartwigsen et~al.(1998)Hartwigsen, Goedecker, and Hutter]{HarGH98}
C.~Hartwigsen, S.~Goedecker, and J.~Hutter.
\newblock Relativistic separable dual-space {G}aussian pseudopotentials from
  {H} to {Rn}.
\newblock \emph{Phys. Rev. B}, 58\penalty0 (7):\penalty0 3641--3662, 1998.

\bibitem[Rappoport and Furche(2010)]{Rappoport:POG:2010}
D.~Rappoport and F.~Furche.
\newblock Property-optimized gaussian basis sets for molecular response
  calculations.
\newblock \emph{J. Chem.\ Phys.}, 133:\penalty0 134105, 2010.

\bibitem[Cai et~al.(2013)Cai, Bai, Pask, and Sukumar]{Cai:2013:HPI}
Y.~Cai, Z.~Bai, J.~E. Pask, and N.~Sukumar.
\newblock Hybrid preconditioning for iterative diagonalization of
  ill-conditioned generalized eigenvalue problems in electronic structure
  calculations.
\newblock \emph{J. Comp.\ Phys.}, 255:\penalty0 16--30, 2013.

\bibitem[Birch(1947)]{Bir47}
F.~Birch.
\newblock Finite elastic strain of cubic crystals.
\newblock \emph{Phys. Rev.}, 71\penalty0 (11):\penalty0 809, 1947.

\bibitem[Schweitzer(2011)]{Schweitzer:2011:SEL}
M.~A. Schweitzer.
\newblock Stable enrichment and local preconditioning in the particle-partition
  of unity method.
\newblock \emph{Numer.\ Math.}, 118:\penalty0 137--170, 2011.

\bibitem[Schweitzer(2013)]{Schweitzer:2013:VML}
M.~A. Schweitzer.
\newblock Variational mass lumping in the partition of unity method.
\newblock \emph{SIAM J. Sci.\ Comput.}, 35\penalty0 (2):\penalty0 A1073--A1097,
  2013.

\bibitem[Lin et~al.(2012)Lin, Lu, Ying, and E]{Lin:2012:ALB}
L.~Lin, J.~Lu, L.~Ying, and W.~E.
\newblock Adaptive local basis set for {Kohn-Sham} density functional theory in
  a discontinuous {Galerkin} framework {I}: {Total} energy calculation.
\newblock \emph{J. Comp. Phys.}, 231\penalty0 (4):\penalty0 2140--2154, 2012.

\end{thebibliography}

\end{document}